\documentclass{elsart}
\journal{Physics Letters B}
\usepackage{graphics}
\usepackage{graphicx}
\usepackage{epsfig}
\usepackage{amssymb}
\usepackage{amsmath}
\usepackage{textcomp}
\usepackage{dcolumn}
\usepackage{array}
\usepackage{varioref}
\usepackage{afterpage}
\usepackage{longtable}

\begin{document}

\begin{frontmatter}

\begin{flushleft}
\hspace*{9cm}BELLE Preprint 2011-2 \\
\hspace*{9cm}KEK Preprint 2010-50 \\
\end{flushleft}

\title{Study of $B^{\pm}\to K^{\pm}(K_SK\pi)^0$ Decay
and Determination of $\eta_c$ and $\eta_c(2S)$ Parameters}

\collab{Belle Collaboration}
  \author[BINP,Novosibirsk]{A.~Vinokurova}, 
  \author[BINP,Novosibirsk]{A.~Kuzmin}, 
  \author[BINP,Novosibirsk]{S.~Eidelman}, 
  \author[BINP,Novosibirsk]{K.~Arinstein}, 
  \author[BINP,Novosibirsk]{V.~Aulchenko}, 
  \author[Lausanne,ITEP]{T.~Aushev}, 
  \author[Sydney]{A.~M.~Bakich}, 
  \author[ITEP]{V.~Balagura}, 
  \author[Melbourne]{E.~Barberio}, 
  \author[Protvino]{K.~Belous}, 
% ...

\address[BINP]{Budker Institute of Nuclear Physics, Novosibirsk, Russian Federation}
% ...
\address[Protvino]{Institute for High Energy Physics, Protvino, Russian Federation}
% ...
\address[ITEP]{Institute for Theoretical and Experimental Physics, Moscow, Russian Federation}
% ...
\address[Lausanne]{\'Ecole Polytechnique F\'ed\'erale de Lausanne, EPFL, Lausanne, Switzerland}
% ...
\address[Melbourne]{University of Melbourne, Victoria, Australia}
% ...
\address[Novosibirsk]{Novosibirsk State University, Novosibirsk, Russian Federation}
% ...
\address[Sydney]{School of Physics, University of Sydney, NSW 2006, Australia}
% ...

\begin{abstract}
We report the results of a study of $B^{\pm}\to K^{\pm}\eta_c$ and $B^{\pm}\to
K^{\pm}\eta_c(2S)$ decays followed by $\eta_c$ and $\eta_c(2S)$ decays
to $(K_SK\pi)^0$. The results are obtained from a data sample 
containing 535 million $B\bar{B}$-meson pairs collected by the Belle
experiment at the KEKB $e^+e^-$ collider. 
We measure the products of the branching fractions 
${\mathcal B}(B^{\pm}\to K^{\pm}\eta_c){\mathcal B}(\eta_c\to
 K_S K^{\pm}\pi^{\mp})=(26.7\pm 1.4(stat)^{+2.9}_{-2.6}(syst)\pm 4.9(model))\times 
10^{-6}$ and ${\mathcal B}(B^{\pm}\to K^{\pm}\eta_c(2S)){\mathcal B}(\eta_c(2S)\to 
K_S K^{\pm}\pi^{\mp})=(3.4^{+2.2}_{-1.5}(stat+model)^{+0.5}_{-0.4}(syst))\times 10^{-6}$.
Interference with the non-resonant component leads to significant
model uncertainty in the measurement of these product branching fractions. Our
analysis accounts for this interference and allows the model
uncertainty to be reduced.
We also obtain the following charmonia masses and widths: 
$M(\eta_c)=(2985.4\pm 1.5(stat)^{+0.5}_{-2.0}(syst))$ 
MeV/$c^2$, 
$\Gamma(\eta_c)=(35.1\pm 3.1(stat)^{+1.0}_{-1.6}(syst))$ 
MeV/$c^2$, $M(\eta_c(2S))=(3636.1^{+3.9}_{-4.2}(stat+model)
^{+0.7}_{-2.0}(syst))$ MeV/$c^2$, $\Gamma(\eta_c(2S))=(6.6^{+8.4}_{-5.1}(stat+model)
^{+2.6}_{-0.9}(syst))$ MeV/$c^2$.
\end{abstract}


\end{frontmatter}

\section{Introduction}

Charmonium states consist of a heavy charm-anticharm quark pair, which
allows the prediction of some of the parameters of these states using 
non-relativistic and relativistic potential 
models~\cite{potential}, lattice QCD~\cite{lattice}, non-relativistic effective 
field theory (NRQCD)~\cite{NRQCD}, and sum rules~\cite{sum} (see the
recent review in~\cite{qwg}). 
The comparison of these predictions
with experimental results provides an opportunity to tune the parameters of
theoretical models and, therefore, improve the accuracy of other values 
predicted by these models.
We have to measure the 
charmonium masses, widths, and product branching fractions with enough accuracy 
to compare them with theoretical predictions. 
Parameters of ($c\bar{c}$) states such as the $\eta_c$ and $\eta_c(2S)$ mesons 
have been studied in
various experiments using a variety of decay channels~\cite{PDG}. 
Tables~\ref{tab:comp1} and~\ref{tab:comp2}, which list a selection of most precise 
mass and width determinations, show that there is quite a large spread of
measured masses and widths of the $\eta_c$ and, especially, $\eta_c(2S)$ mesons
resulting in large scale factors for the world average values~\cite{PDG}.
Moreover, our knowledge of hadronic decays of these charmonia is rather poor.

In the Belle and BaBar $B$ factory experiments, 
charmonia are produced in various ways: from fragmentation in electron-positron
annihilation,
from two-photon processes, and in $B$ decays. The advantages of
$B^{\pm}\to K^{\pm}c\bar{c}$ decay are the relatively large reconstruction 
efficiency, small background,
and the fixed quantum numbers ($J^P=0^-$) of the initial state.
Here we consider the following decays of charged $B$ 
mesons:
\begin{center}
$B^{\pm}\to K^{\pm}\eta_c\to K^{\pm}(K_SK\pi)^0$,\\
$B^{\pm}\to K^{\pm}\eta_c(2S)\to K^{\pm}(K_SK\pi)^0$.
\end{center}
A 492 fb$^{-1}$ data sample provides an
opportunity to determine the corresponding products of the branching fractions 
as well as masses 
and widths of the $\eta_c$ and $\eta_c(2S)$ mesons.
\begin{table}[!h]
\begin{center}
\caption{Previously measured $\eta_c$ parameters.}
\begin{tabular}{|c||c|c|c|} 
\hline
Experiment & Process & Mass, MeV/$c^2$ & Width, MeV/$c^2$\\
\hline\hline
BaBar \cite{Babar10} & $\gamma\gamma\to\eta_c\to K_S^0K^{\pm}\pi^{\mp}$ &
$2982.2\pm 0.4\pm 1.6$ & $31.7\pm 1.2\pm 0.8$\\
Belle \cite{Belle08_2} & $\gamma\gamma\to\eta_c\to K_SK^{\pm}\pi^{\mp}$ & $2981.4\pm
0.5\pm 0.4$ & $36.6\pm 1.5\pm 2.0$\\
BaBar \cite{Babar08} & $B\to\eta_c K^{(*)}\to K\bar{K}\pi K^{(*)}$ &
$2985.8\pm 1.5\pm 3.1$ & $36.3^{+3.7}_{-3.6}\pm 4.4$\\
Belle \cite{Belle08_1} & $\gamma\gamma\to\eta_c\to$hadrons & $2986.1\pm
1.0\pm 2.5$ & $28.1\pm 3.2\pm 2.2$\\
CLEO \cite{Cleo04} & $\gamma\gamma\to\eta_c\to K^0_S K^{\pm}\pi^{\mp}$
& $2981.8\pm 1.3\pm 1.5$ & $24.8\pm 3.4\pm 3.5$\\
BES \cite{Bes03} & $J/\psi\to\gamma\eta_c$ & $2977.5\pm 1.0\pm 1.2$ &
$17.0\pm 3.7\pm 7.4$\\
E835 \cite{e835} & $p\bar{p} \to \gamma\gamma$ &
$2984.1 \pm 2.1 \pm 1.0$ & $20.4^{+7.7}_{-6.7} \pm 2.0$ \\ 
\hline
\end{tabular}
\label{tab:comp1}
\end{center}
\end{table}
\begin{table}[!h]
\begin{center}
\caption{Previously measured $\eta_c(2S)$ parameters.}
\begin{tabular}{|c||c|c|c|} 
\hline
Experiment & Process & Mass, MeV/$c^2$ & Width, MeV/$c^2$\\
\hline\hline
Belle \cite{Belle08_2} & $\gamma\gamma\to\eta_c(2S)\to K_SK^{\pm}\pi^{\mp}$ &
 $3633.7\pm 2.3\pm 1.9$ & $19.1\pm 6.9\pm 6.0$\\
Belle \cite{Belle07} & $e^+e^-\to J/\psi c\bar{c}$ &
$3626\pm 5\pm 6$ & ---\\
BaBar \cite{Babar05} & $e^+e^-\to J/\psi c\bar{c}$ & $3645.0\pm
5.5^{+4.9}_{-7.8}$ & $22\pm 14$\\
CLEO \cite{Cleo04} & $\gamma\gamma\to\eta_c(2S)\to K^0_S K^{\pm}\pi^{\mp}$
& $3642.9\pm 3.1\pm 1.5$ & $6.3\pm 12.4\pm 4.0$\\
BaBar \cite{Babar04} & $\gamma\gamma\to\eta_c(2S)\to K\bar{K}\pi$ &
$3630.8\pm 3.4\pm 1.0$ & $17.0\pm 8.3\pm 2.5$\\
Belle \cite{Belle02} & $B\to KK_S K^{\pm} \pi^{\mp}$ & $3654\pm 6\pm 8$ & $<55$\\
\hline
\end{tabular}
\label{tab:comp2}
\end{center}
\end{table}

At all stages of this analysis we consistently take into account the interference 
between the $B^{\pm}\to K^{\pm}\eta_c$ and $B^{\pm}\to K^{\pm}\eta_c(2S)$
decays and the decay $B^{\pm}\to K^{\pm}(K_S K\pi)^0$, which has the same
final state but no intermediate charmonium particle.

\section{Event selection}

The results are based on a data sample that contains $535\times 10^6$
$B\bar{B}$ pairs, collected with the Belle detector at the KEKB 
asymmetric-energy $e^+e^-$ collider~\cite{KEKB} operating at the $\Upsilon(4S)$
resonance.

  The Belle detector~\cite{Belle} is a large-solid-angle magnetic spectrometer
that consists
of a silicon vertex detector (SVD), 
a 50-layer central drift chamber
(CDC) for charged particle tracking and specific ionization measurement 
($dE/dx$), an array of aerogel threshold Cherenkov counters (ACC), 
time-of-flight scintillation counters (TOF), and an array of 8736 CsI(Tl) 
crystals for electromagnetic calorimetry (ECL) located inside a superconducting
solenoid coil that provides a 1.5~T magnetic field. An iron flux return located
outside the coil is instrumented to detect $K_L^0$ mesons and identify muons
(KLM). 
We use a GEANT-based Monte Carlo (MC) simulation to
model the response of the detector and determine its acceptance~\cite{sim}.

Pions and kaons are separated by combining the responses of 
the ACC and the TOF with $dE/dx$ measurements in the CDC
 to form a likelihood $\mathcal{L}$($h$) where $h=\pi$ or $K$. 
Charged particles 
are identified as pions or kaons using the likelihood ratio 
$\mathcal{R}$:
\[{\rm {\mathcal {R}}}(K)=\frac{{\mathcal{L}}(K)}{{\mathcal{L}}(K)+{\mathcal{L}}(\pi)};~~
{\rm {\mathcal R}}(\pi)=
\frac{{\mathcal{L}}(\pi)}{{\mathcal{L}}(K)+{\mathcal{L}}(\pi)}=1-{\rm {\mathcal R}}(K).\]

Charged tracks are selected with requirements based on the $\chi^2$ of the
track fits and the impact parameters relative to the interaction point. We
require that the polar angle of each track be in the angular range
18$^{\circ}$ -- 152$^{\circ}$
and that the track momentum perpendicular to the positron beamline be greater 
than 100 MeV/$c$.

Charged kaon candidates are identified by the requirement 
${\rm {\mathcal R}}(K)>0.6$, 
which has an efficiency of $90\%$ and a pion misidentification probability of
3 -- 10\% depending on momentum.
For pion candidates we require \mbox{${\rm {\mathcal R}}(\pi)>0.2$}.
$K_S$ candidates are reconstructed via the $\pi^+\pi^-$ mode. We apply the 
following cut
on the $\pi^+\pi^-$ invariant mass: $0.489$ GeV/$c^2$ $<M(\pi^+\pi^-)<0.505$ 
GeV/$c^2$. 
The flight length of the $K_S$ is required to lie within the interval 
$[0.1:20]$ mm. The condition on the $K_S$ direction angle $\varphi$ is
$cos(\varphi)>0.95$.

$B$ meson candidates are identified 
by their center-of-mass (c.m.)\ energy difference 
$\Delta E=(\sum_iE_i)-E_{\rm b}$, and the
beam-constrained mass 
$M_{\rm bc}=\sqrt{E^2_{\rm b}-(\sum_i\vec{p}_i)^2}$, where 
$E_{\rm b}=\sqrt{s}/2$ is the beam energy in the $\Upsilon(4S)$
c.m.\ frame, and $\vec{p}_i$ and $E_i$ are the c.m.\ three-momenta and 
energies, respectively, of the $B$ meson candidate decay products. The 
signal region is defined as: $|{\Delta}E|<0.03$ GeV, 
$5.273$ GeV/$c^2$ $<M_{\rm bc}<5.285$ GeV/$c^2$. The ${\Delta}E$ sideband region 
is defined as $||{\Delta}E|-0.06$(GeV)$|<0.03$ GeV.

To suppress the large continuum background ($e^+e^-\to q\bar{q}$,
where $q=u,d,s,c$), topological variables are used. Since  
the produced $B$ mesons 
are nearly at rest in the c.m.\ frame, the signal
tends to be 
isotropic while continuum $q\bar{q}$ background tends
to have  a two-jet structure. We use the angle between the thrust axis of 
the $B$ candidate and that of the rest of the event ($\Theta_{\rm thrust}$)
to discriminate between these two cases. The distribution of
$|\cos\Theta_{\rm thrust}|$ is strongly peaked near $|\cos\Theta_{\rm thrust}|=1$
for $q\bar{q}$ events and is nearly uniform for  $\Upsilon(4S)\to
B\bar{B}$ events.
We require $|\cos\Theta_{\rm thrust}|<0.8$.

If there is more than one combination that satisfies the selection 
criteria, we select the $B$
candidate with the minimum difference $|M(K_S)-M(\pi^+\pi^-)|$ and the minimum 
difference between the vertex z-coordinates of the kaon and pion
from the charmonium decay, and the kaon from the $B$ decay. 
If the final state includes two kaons of the same
charge, as in $K^{\pm}(K^{\pm}K_S\pi^{\mp})$, we must choose the one
from the charmonium decay. For this purpose we select the candidate with
the minimum 
difference $|M(\eta_c/\eta_c(2S))-M(K_SK^{\pm}\pi^{\mp})|$. The mean number of 
multiple candidates per event is 1.6.

\section{Interference study}

The data sample can contain signal, which has the same final state
as a resonant decay but does not include a charmonium resonance. The contribution 
of these events is referred to as the non-resonant amplitude. 
Since the final state is the
same, this amplitude interferes with the signal. However, if 
the final particles 
form narrow resonances such as $D$,
$D_S$, and $\phi$ mesons, the interference effect cancels after
integration over mass. Thus, we can 
reject some background channels by applying appropriate cuts on the mass combinations of 
the final state particles. If the intermediate resonances 
have a substantial width or the $B$
meson decays directly into the final state particles, the effect of interference
must be taken into account.

\begin{figure}[!h]
\begin{center}
\begin{tabular}{cc}
\includegraphics[height=5 cm]{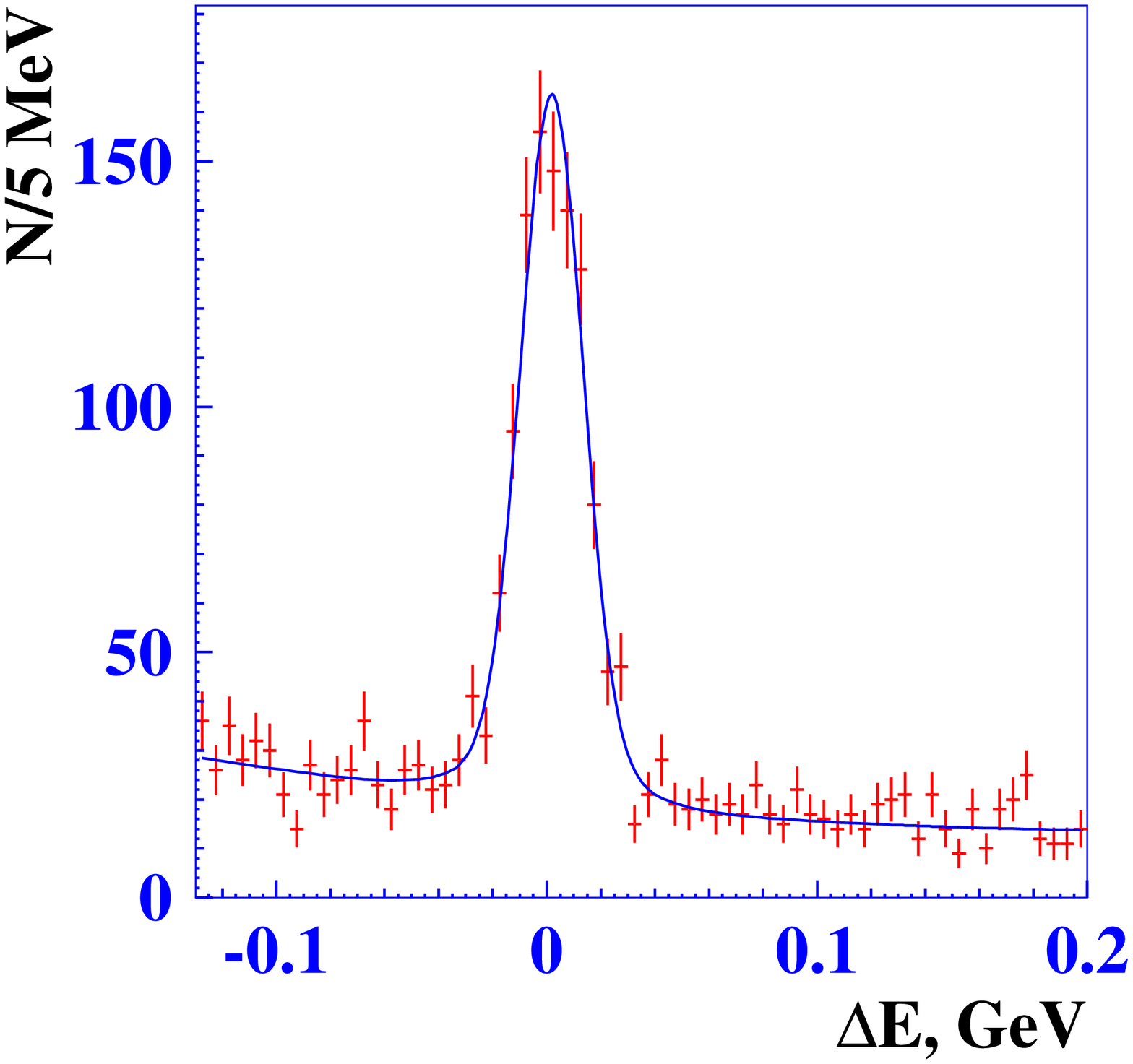}&
\includegraphics[height=5 cm]{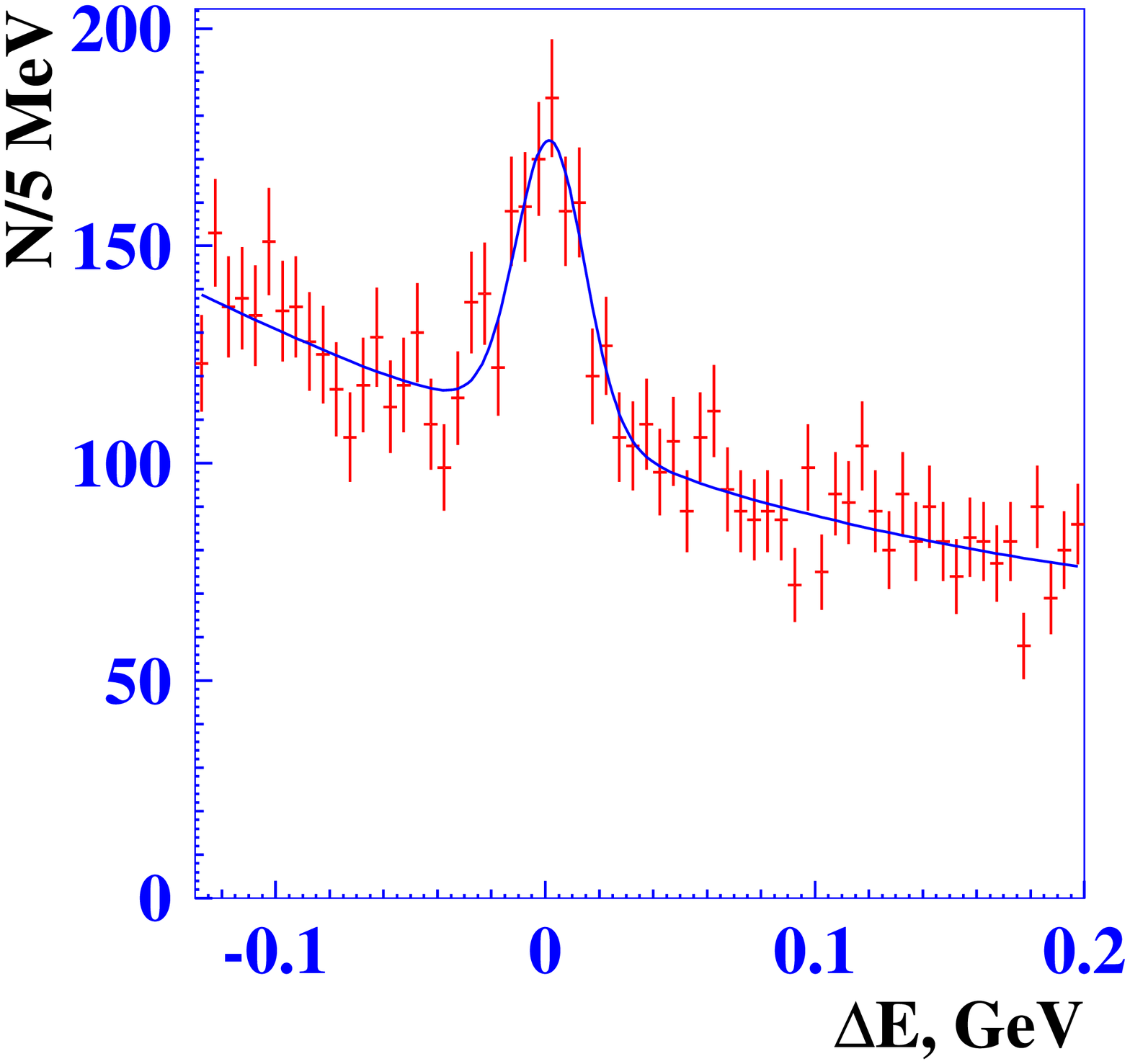}\\
\end{tabular}
\caption{$\Delta E$ distributions in the signal (left) and
  sideband (right) regions for $B^{\pm}\to K^{\pm}\eta_c\to K^{\pm}
(K_SK\pi)^0$ candidates.}
\label{pic:de1}
\end{center}
\end{figure}

\begin{figure}[!h]
\begin{center}
\begin{tabular}{cc}
\includegraphics[height=5 cm]{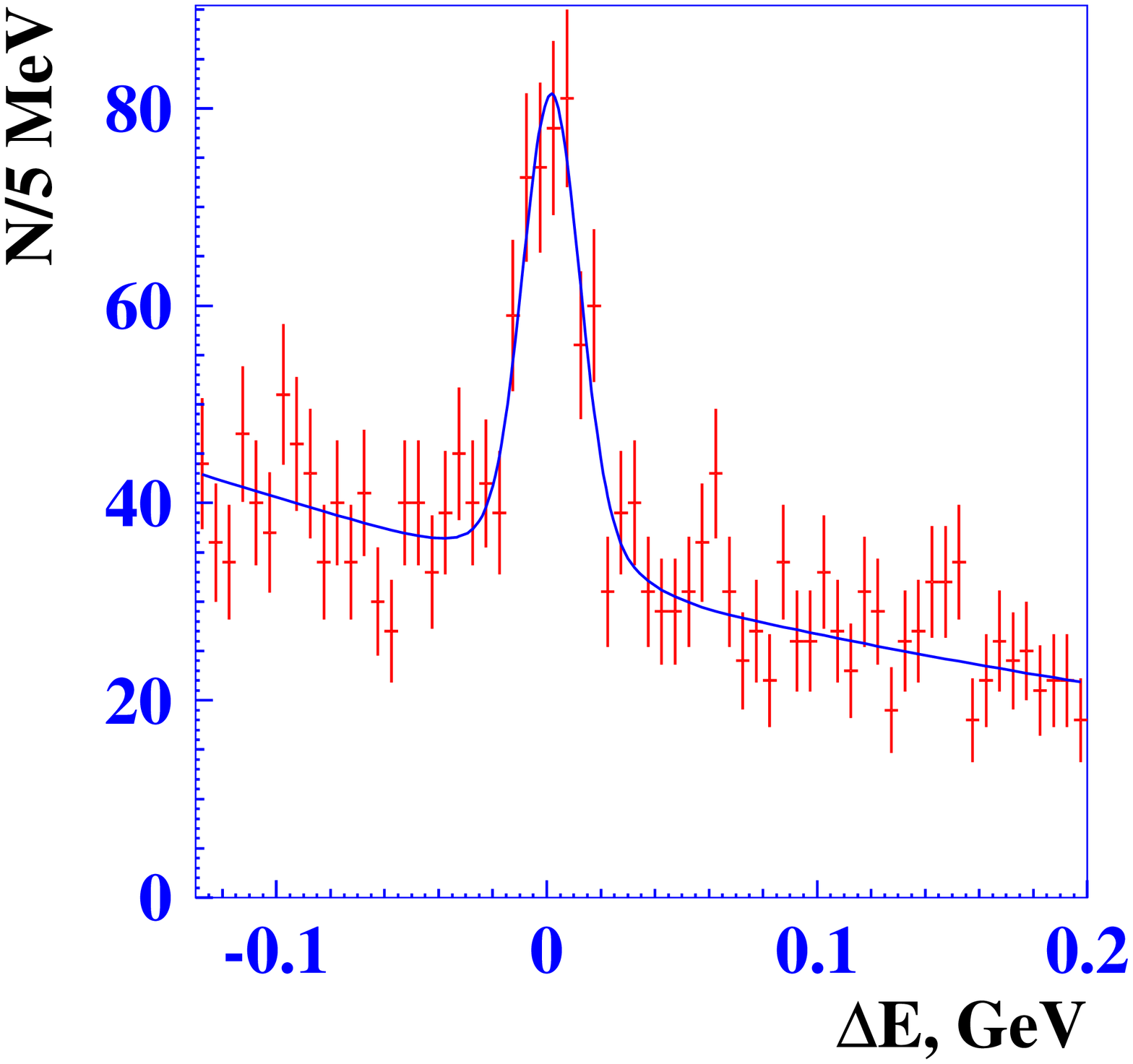}&
\includegraphics[height=5 cm]{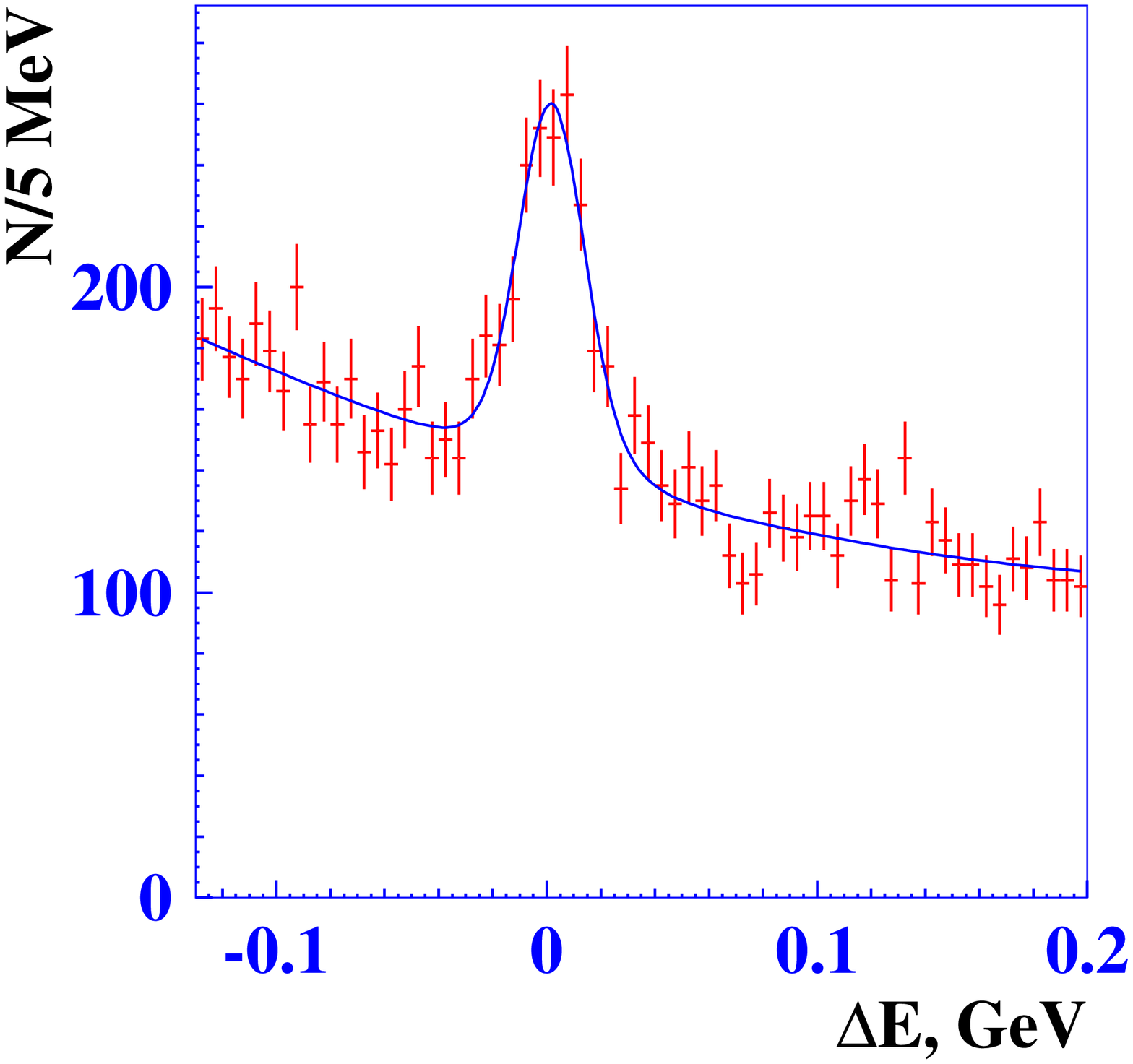}\\
\end{tabular}
\caption{$\Delta E$ distributions in the signal (left) and
  sideband (right) regions for $B^{\pm}\to K^{\pm}\eta_c(2S)\to K^{\pm}
(K_SK\pi)^0$ candidates.}
\label{pic:de2}
\end{center}
\end{figure}

The non-resonant contribution can be seen as a peak in the $\Delta E$ 
distribution in
the charmonium sideband regions shown in Figs.~\ref{pic:de1} 
and~\ref{pic:de2} (plots on the right)\footnote{The $\eta_c$ meson mass signal region is ($2.92$ -- $3.04$) 
GeV/$c^2$, the sideband is ($2.54$ -- $2.86$) GeV/$c^2$ and ($3.14$
  -- $3.46$) GeV/$c^2$. The $\eta_c(2S)$
  meson mass signal region is ($3.58$ -- $3.7$) GeV/$c^2$, the sideband is
 ($3.14$ -- $3.46$) GeV/$c^2$ and ($3.72$ -- $4$) GeV/$c^2$.}.
We fit the $\Delta E$ distributions with 
the sum of a Gaussian distribution 
and a second-order polynomial function. From these fits we obtain the number of 
events in the signal region
$N_{obs}$ and in the sideband region $N_{sb}$, which can be rescaled to obtain
the number of non-resonant events $N_{non-res}$. 
In the $\eta_c$ case
$N_{obs}=889\pm 37(stat)$ and $N_{non-res}=87\pm 11(stat)$. In
the $\eta_c(2S)$ case $N_{obs}=279\pm 29(stat)$ and
$N_{non-res}=156\pm 13(stat)$. 

Different values of the interference phase can give significant variations in
the number of signal events while the total number of
observed events remains the same. Using the hypotheses of maximal constructive and 
destructive 
interference, we would obtain $410$ and $1550$ signal events, respectively.
Therefore, the model uncertainty in the number of $\eta_c$ signal events
is rather large: $N_{signal}=980\pm
570(model)$. It would be even larger for the $\eta_c(2S)$ decay.
A dedicated study of the interference effect allows this
uncertainty to be reduced.

\begin{figure}[!h]
\begin{center}
\includegraphics[height=4 cm]{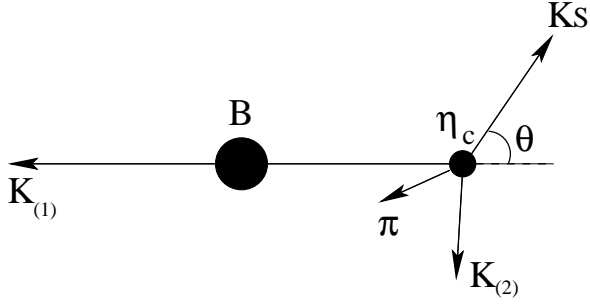}
\caption{The decay $B^{\pm}\to K^{\pm}\eta_c\to K^{\pm}(K_S K\pi)^0$.}
\label{pic:7}
\end{center}
\end{figure}

In the $B\to K_{(1)} K_S K_{(2)}\pi$ decay (see Fig.~\ref{pic:7}) there are
four particles in the final state, which gives $4\times 3$ measured
parameters. Taking into account the four constraints of
energy-momentum conservation and integrating over the three angles that
characterize the $B$ decay (it is
a pseudoscalar and there should be no dependence on these angles), we have 5 
independent variables to describe the amplitude of the process.
We chose the following parameters:
$K_SK_{(2)}\pi$ invariant mass, two Dalitz variables for the $\eta_c$ (or $\eta_c(2S)$)
decay -- $q_1^2$ and $q_2^2$ (for example, $M(K_{(2)}\pi)^2$ and
$M(K_S\pi)^2$), the angle between the $K_S$ and $K_{(1)}$ in the 
rest frame of the $K_{(2)}K_S\pi$ system ($\theta$), and the angle
between the planes of the $K_{(1)}-\pi$ and $K_{(1)}-K_S$ in the same system ($\phi$).

\begin{figure}[!h]
\begin{center}
\includegraphics[height=8 cm]{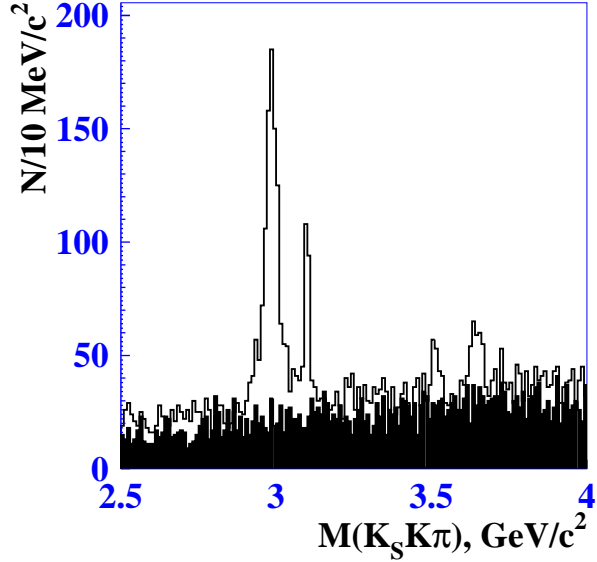}
\caption{The signal distribution of ($K_S
  K^{\pm}\pi^{\mp}$) invariant mass in the $B^{\pm}\to
  K^{\pm}(K_S K\pi)^0$ decay. The charmonium states $\eta_c$,
  $J/\psi$, $\chi_{c1}$, and $\eta_c(2S)$ (in order of mass)
  can be seen. The solid histogram is the combinatorial
  background determined from the $\Delta E$ sideband region.}
\label{pic:charm}
\end{center}
\end{figure}

The $M(K_S K\pi)$ distribution has four peaks corresponding to
$\eta_c$, $J/\psi$, $\chi_{c1}$ and $\eta_c(2S)$ production (see
Fig.~\ref{pic:charm}). 
In addition to these peaks, there is a non-resonant signal, which interferes
with the $\eta_c$ (or $\eta_c(2S)$) signal\footnote{We assume
that the non-resonant component is described
by a smooth function.}. Unfortunately,
the shape of the one dimensional (1-D) mass distribution alone does
not allow the
interference contribution to be obtained, so other variables should be used.

\begin{figure}[!h]
\begin{center}
\begin{tabular}{cc}
\includegraphics[height=5 cm]{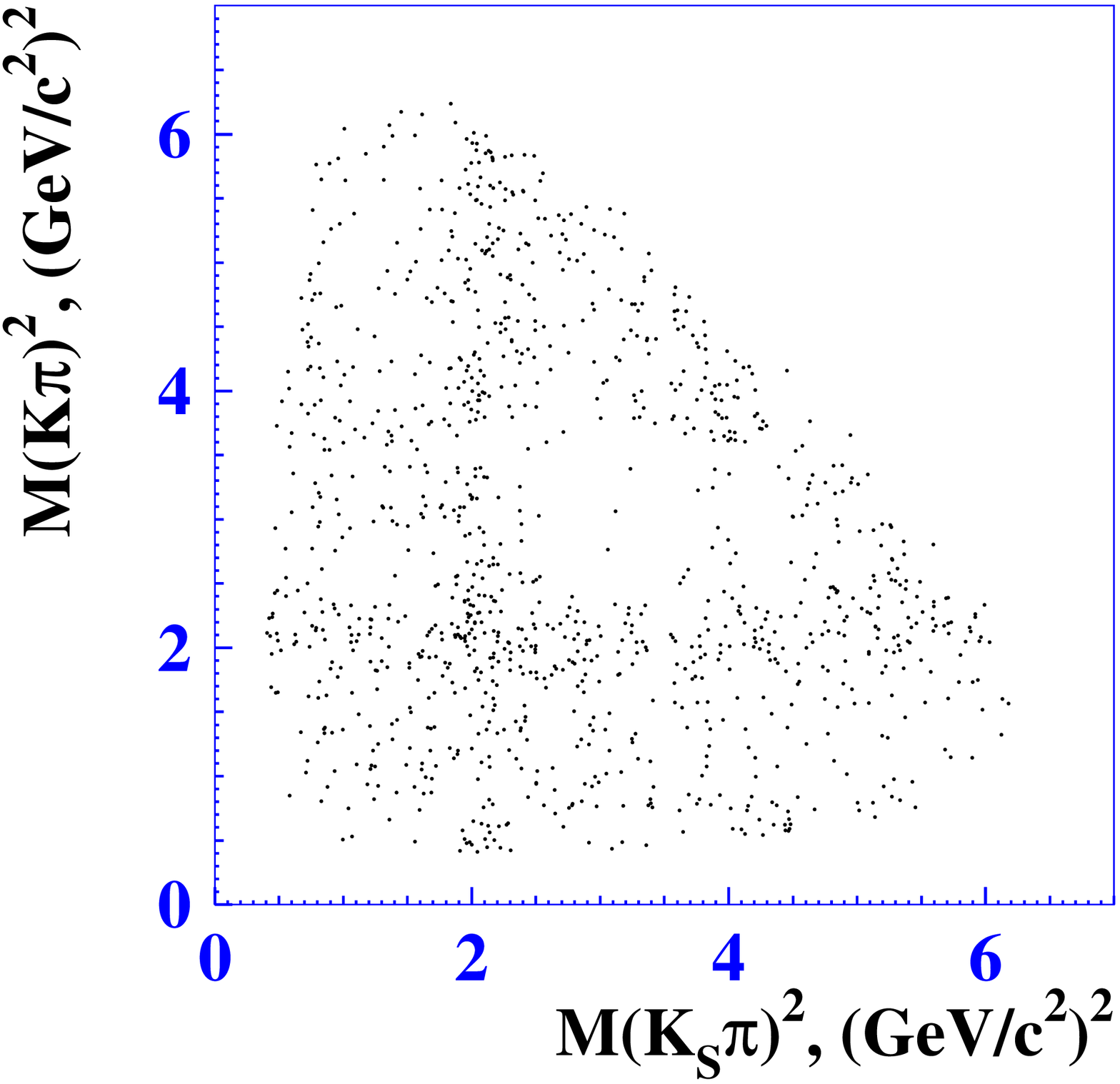}&
\includegraphics[height=5 cm]{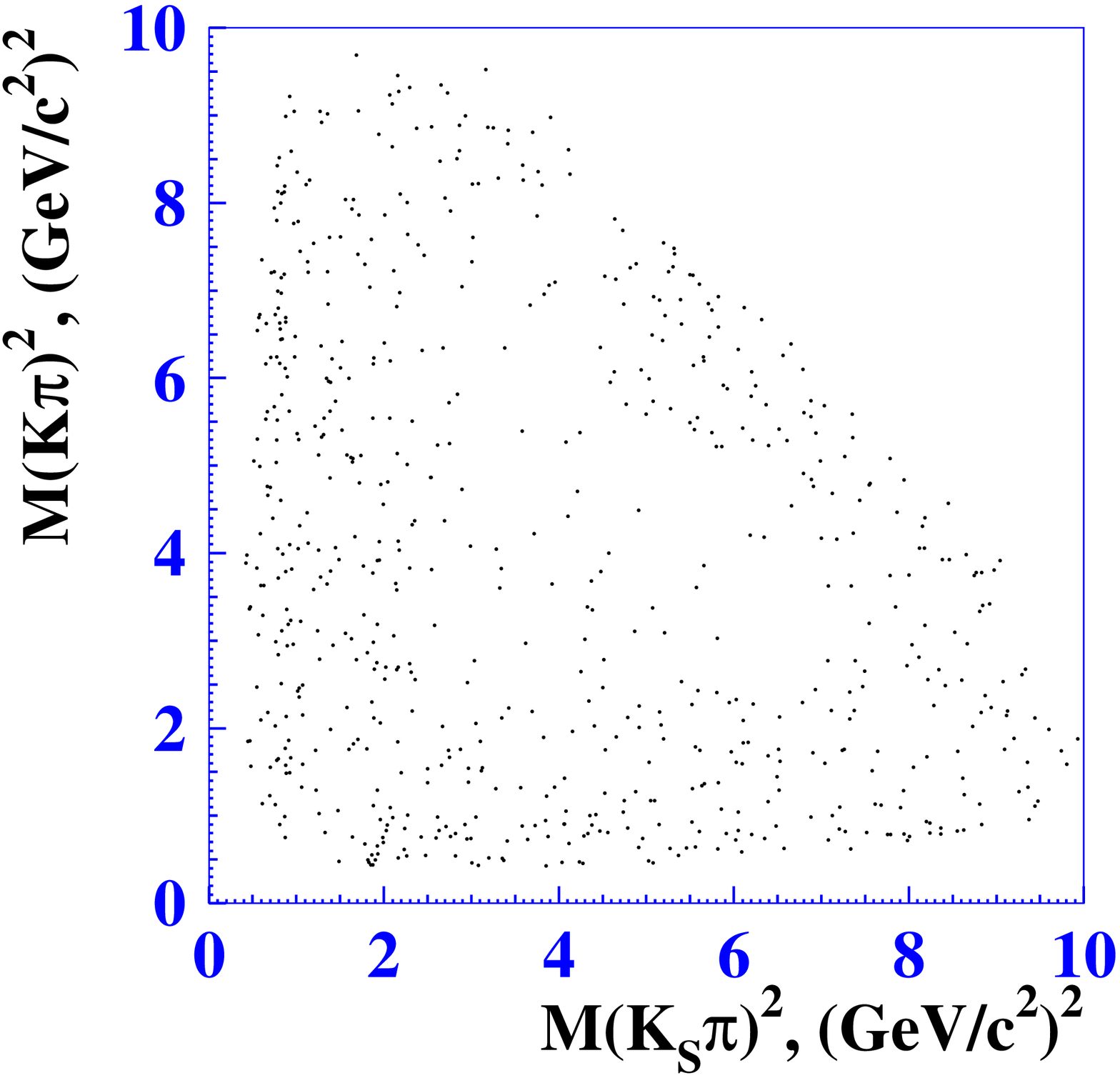}\\
\end{tabular}
\caption{The Dalitz distributions in the $\eta_c$ (left) and $\eta_c(2S)$ 
(right) signal regions.}
\label{pic:1}
\end{center}
\end{figure}

The Dalitz plots of 3-body $\eta_c$ and $\eta_c(2S)$ decays are shown in 
Fig.~\ref{pic:1}. In
the $\eta_c$ case the 
distribution is not uniform and has a peaking structure around a $K\pi$
mass of 1.4 GeV/$c^2$, which can be a combination of the $K^*(1410)$,
$K_0^*(1430)$, and $K_2^*(1430)$ states. However, the statistics are 
rather low, so it is impossible
to determine with acceptable accuracy either the mass and width of these
states, or the product branching fractions of $\eta_c$ decay modes 
involving these states. The small number of events does not
allow the Dalitz analysis to be efficiently performed and makes it difficult to use the Dalitz 
plot variables 
to distinguish the $\eta_c$ signal and
non-resonant amplitudes. The same conclusion holds for the
$\eta_c(2S)$ decay.

\begin{figure}[!h]
\begin{center}
\begin{tabular}{cc}
\includegraphics[height=5 cm]{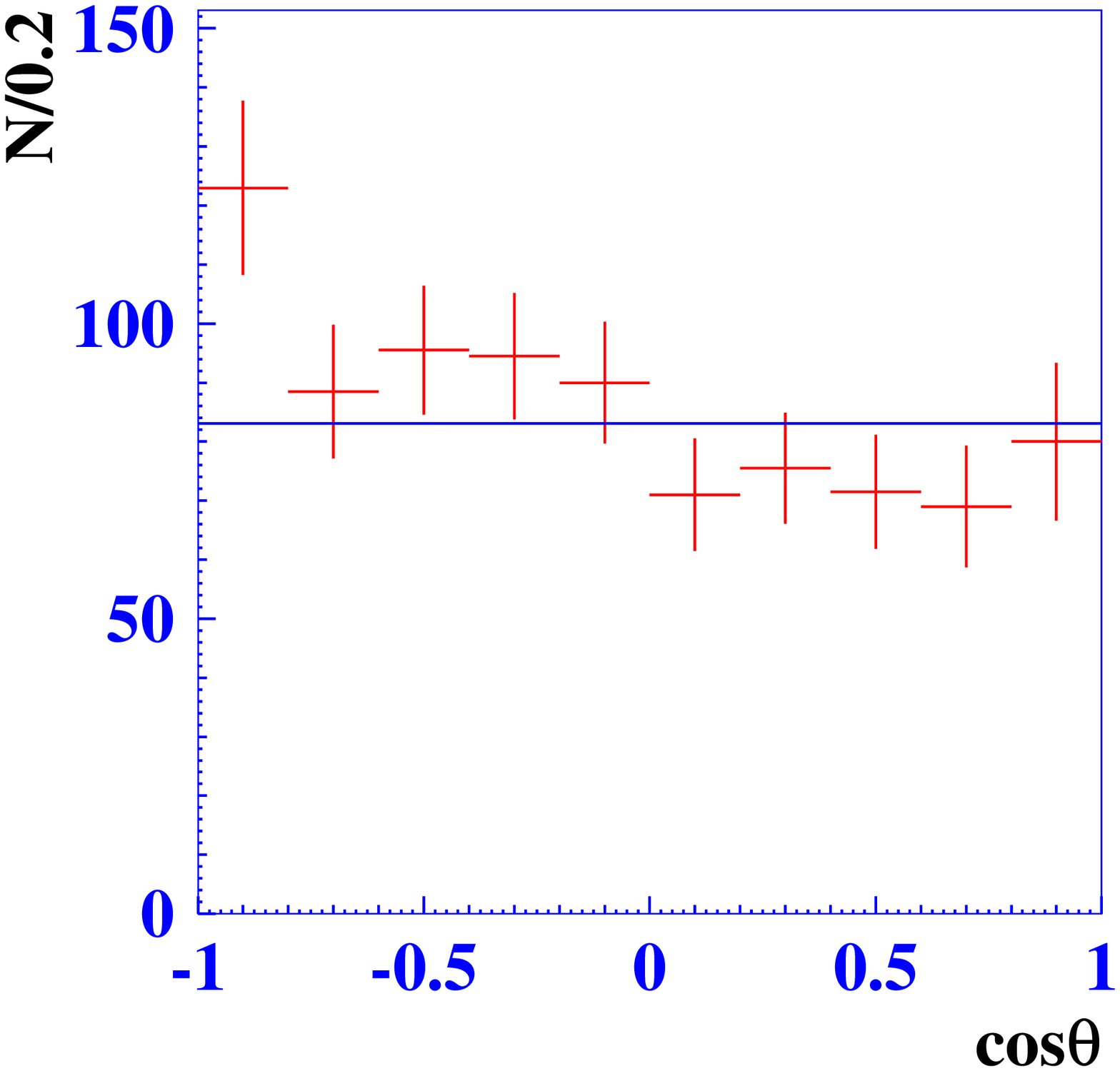}&
\includegraphics[height=5 cm]{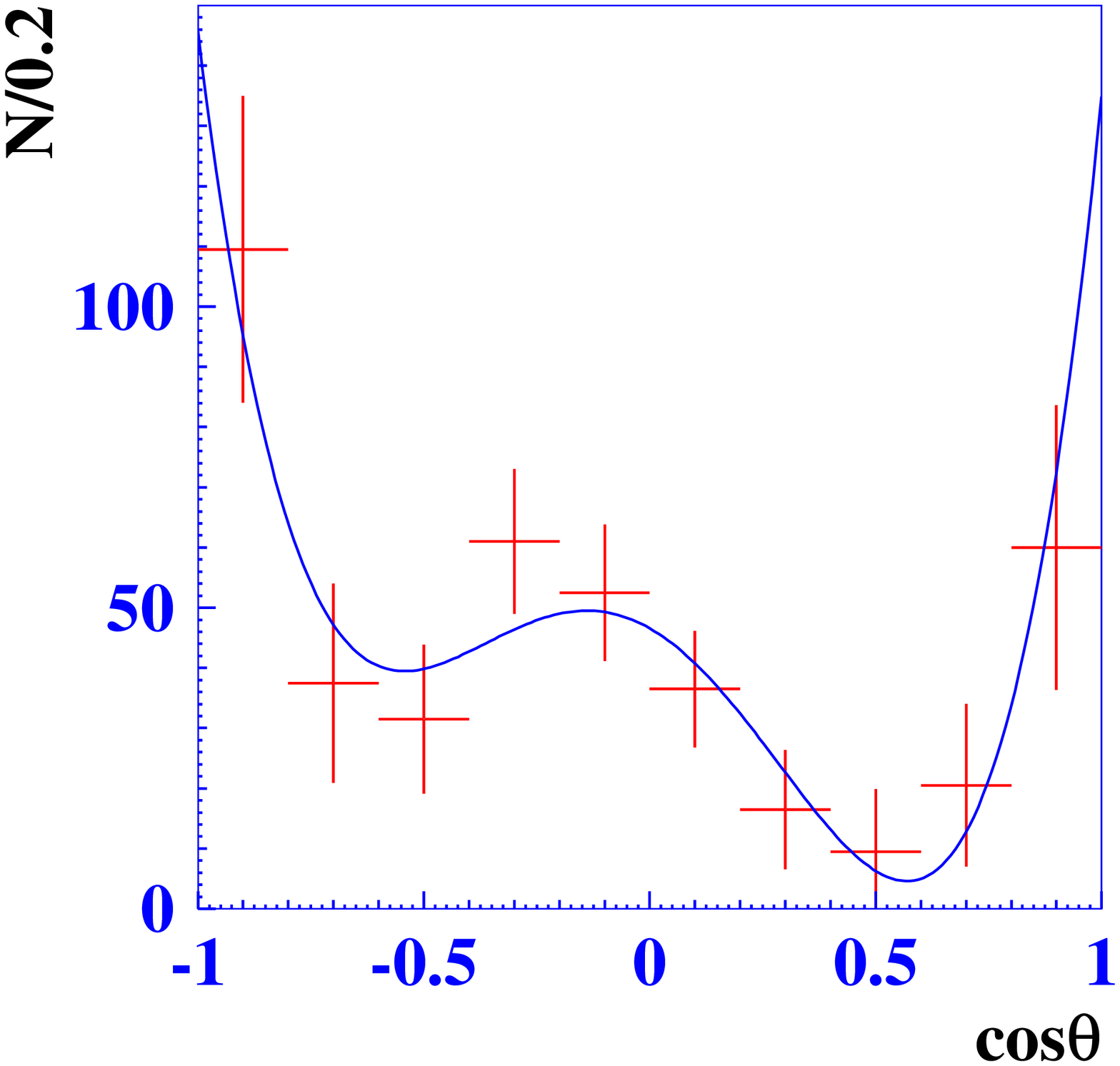}\\
\end{tabular}
\caption{Approximation of the $\cos\theta$ distribution in the $\eta_c$ 
signal region by
an S-wave (left) 
and in the sideband region by a sum of S-, P-, and D-waves (right). The combinatorial 
background is subtracted.}
\label{pic:cos1}
\end{center}
\end{figure}

\begin{figure}[!h]
\begin{center}
\begin{tabular}{cc}
\includegraphics[height=5 cm]{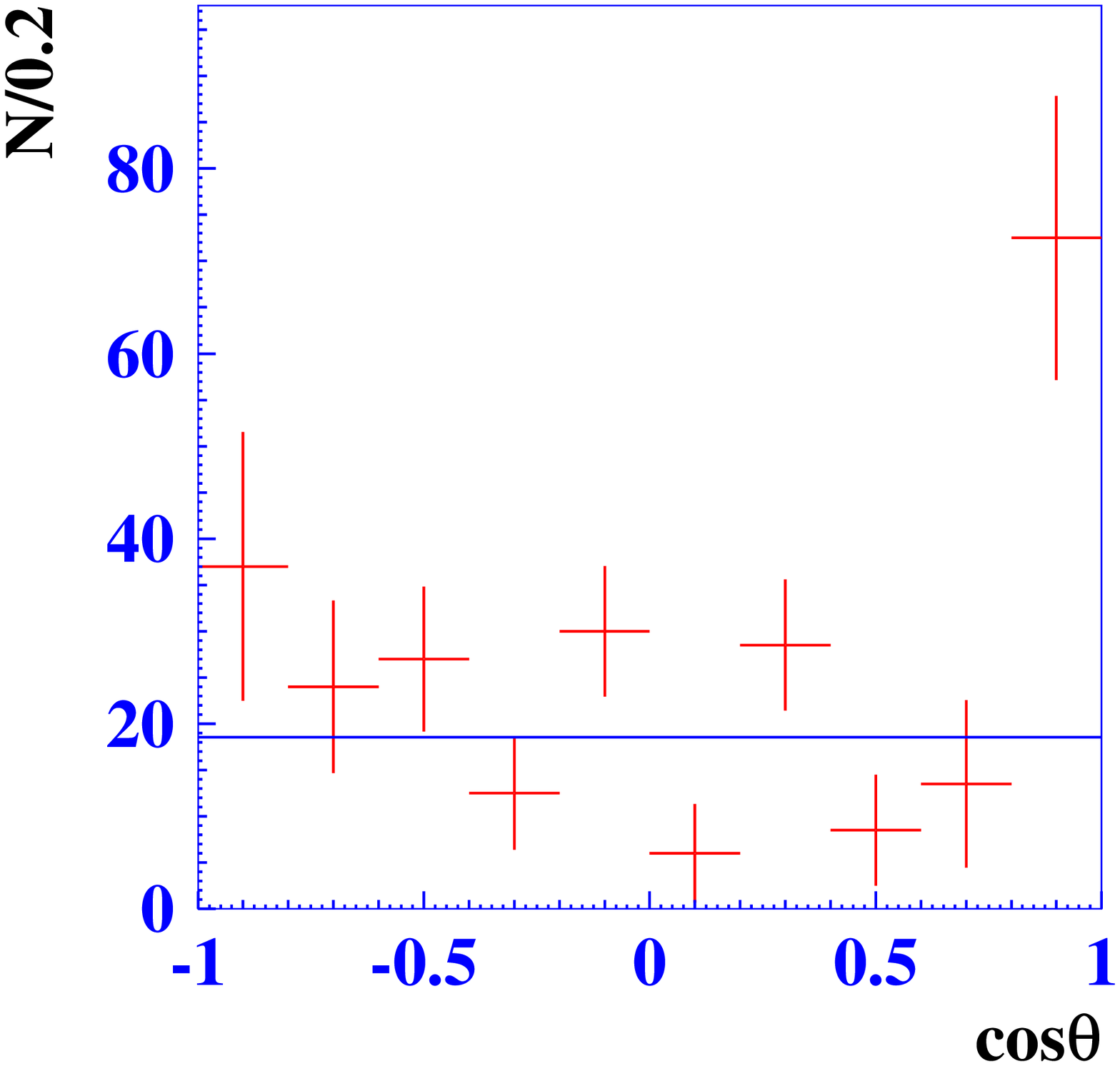}&
\includegraphics[height=5 cm]{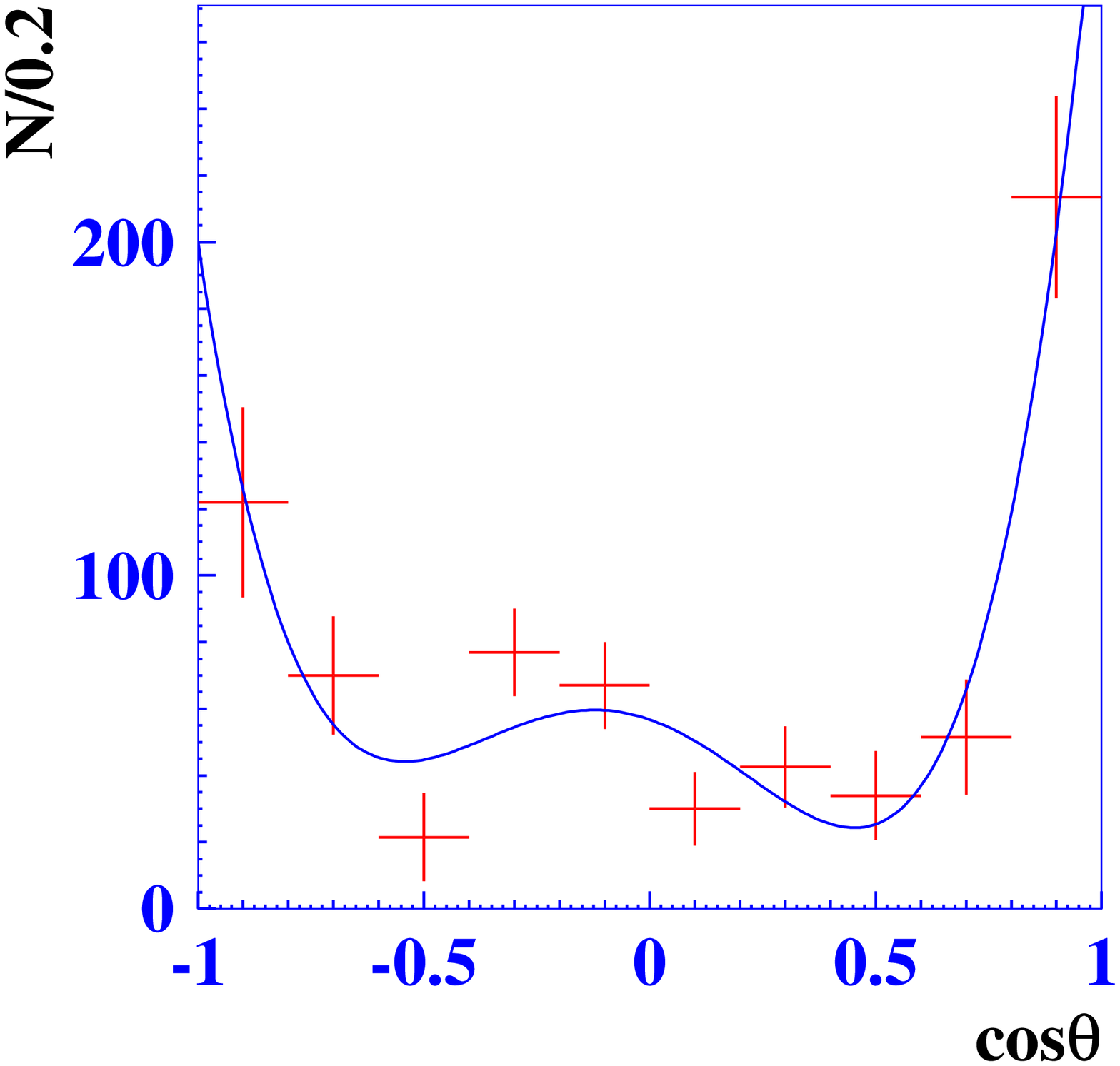}\\
\end{tabular}
\caption{Approximation of the $\cos\theta$ distribution in the 
$\eta_c(2S)$ signal region by an S-wave
(left) and in the sideband region by a sum of S-, P-, and D-waves (right). The 
combinatorial background is subtracted.}
\label{pic:cos2}
\end{center}
\end{figure}

Another variable that can be used for the amplitude separation is
$\cos{\theta}$. 
Since the $\eta_c$ and $\eta_c(2S)$ are pseudoscalars ($J^P=0^-$), we
expect a uniform
distribution in $\cos{\theta}$. In Figs.~\ref{pic:cos1}
and~\ref{pic:cos2}
the $\cos\theta$ distributions are shown for the $\eta_c$ and $\eta_c(2S)$
signal and 
sideband regions, while the combinatorial background is subtracted. 
One can see that the sideband distribution has 
contributions from higher angular waves. A good fit can be obtained with the
sum of S-, 
P-, and D-waves. The signal region also contains         
non-resonant background but mostly consists of signal events, so the S-wave 
contribution here is dominant.
Separation of the P- and D-waves from the S-wave in the
non-resonant background allows the uncertainty from interference
to be reduced.

Thus, we analyze a 2-D $M(KK_S\pi)$--$\cos{\theta}$ histogram assuming 
that the non-resonant signal amplitude is constant within the  
($2.5$ -- $3.46$) and ($3.14$ -- $4.06$) GeV/$c^2$ mass ranges. 
The number of events in a single bin is 
$N_{bin}=N_{\Delta E signal}-k\cdot N_{\Delta E sideband}$, where the coefficient $k$ is
used to normalize the number of events in the $\Delta E$ sideband region. 
We minimize the likelihood function, assuming that the events in the signal
and sideband regions are described by the Poisson statistics.
In the $\eta_c$ analysis the bin size along the
$\cos{\theta}$ axis is $0.2$ (9 bins), while along the $M(K_S K\pi)$ axis
it is $10$ MeV/$c^2$ in
the signal region and $150/130$ MeV/$c^2$ in the left/right sideband region 
(44 bins). 
For the $\eta_c(2S)$ the bin size along the
$\cos{\theta}$ axis is also $0.2$ (9 bins), while along the $M(K_S K\pi)$ axis
it is
$16$ MeV/$c^2$ in the signal region and $130$ MeV/$c^2$ in the sideband region
(29 bins).

We exclude the $J/\psi$ region ($3.07$ --
 $3.13$ GeV/$c^2$) from the fit because the interference
of the $J/\psi$ and non-resonant signal is negligible due to the
small width of the former and inclusion of this region does not constrain
the $\eta_c$ interference contribution. Moreover, the $J/\psi$
angular distribution has contributions from several amplitudes that are not
well determined. The same arguments apply to the exclusion of the
$\chi_{c1}$ mass region ($3.48$ -- $3.54$ GeV/$c^2$). 
We perform separate fits to the $J/\psi$ and $\chi_{c1}$ with Gaussian functions using 
1-D $K_SK\pi$ invariant mass distributions from signal MC and data. After 
comparing the obtained widths we determine the degradation of the resolution
in data. Taking this into account, we recalculate
the detector resolution obtained from 
signal MC in the $\eta_c$ and $\eta_c(2S)$ regions. We obtain 
$\sigma(\eta_c)=(6.2\pm 1.1)$ MeV and $\sigma(\eta_c(2S))=(9.8\pm 1.7)$ MeV.

The fitting function can be represented as the square of the absolute value of the sum
of the signal and non-resonant amplitudes integrated over all
variables except $M(K_SK\pi)$ and $\cos{\theta}$:
\begin{eqnarray}
F(s,x)&=&
\int\int\int\int_{x-\frac{\delta}{2}}^{x+\frac{\delta}{2}}\int_{s-\frac{\Delta}{2}}^{s+\frac{\Delta}{2}}
(1+\varepsilon_1 x'+\varepsilon_2 x'^2)\cdot\nonumber\\
& &\Biggl|\left(\frac{\sqrt{N}}{s'-M^2+iM\Gamma}A_{\eta}(q_1^2,q_2^2)+\alpha
A_S(q_1^2,q_2^2)\right)S(x')+ \nonumber\\ 
& &\beta A_P(q_1^2,q_2^2)P(x')+\gamma
A_D(q_1^2,q_2^2)D(x')\Biggr|^2
ds'dx'dq_1^2dq_2^2d\phi,
\label{eq:f}
\end{eqnarray}
where 
$x=\cos\theta$, $s=M^2(K_SK\pi)$;
$q_1^2$ and $q_2^2$ are Dalitz plot variables;
$\varepsilon_1$ and $\varepsilon_2$ are constants that
  characterize the efficiency dependence on $x$ and are determined from MC;
$\delta$ and $\Delta$ are the bin widths in $\cos{\theta}$
  and $M(K_SK\pi)$ invariant mass, respectively;
$M$ and $\Gamma$ are mass and width of the $\eta_c$ ($\eta_c(2S)$) 
meson;
$N$ is the $\eta_c$ ($\eta_c(2S)$) signal yield; 
$\alpha$, $\beta$, $\gamma$ are the relative fractions of the S-, P-, and
  D-waves, respectively;
$S=\frac{1}{\sqrt{2}}$, $P=\sqrt{\frac{3}{2}}x$,
  $D=\frac{3}{2}\sqrt{\frac{5}{2}}(x^2-\frac{1}{3})$ are the functions
  characterizing the angular dependence of the S-, P-, and D-waves,
  respectively;
$A_{\eta}$ is the signal S-wave amplitude, $A_{S,P,D}$ are the background
  S-, P-, and D-wave amplitudes, respectively. The absolute values of the
  amplitudes squared are normalized to unity:  
\begin{equation}
\int\int\int
|A_{\eta,S,P,D}(q_1^2,q_2^2)|^2dq_1^2dq_2^2d\phi=1.
\end{equation}
To account for the momentum resolution, Eq.~(\ref{eq:f}) is convolved with 
a Gaussian detector resolution function that is determined from the MC and 
calibrated from the $J/\psi$ ($\chi_{c1}$) width in data.

This function is determined by
15 parameters: 
$N$, $M$, $\Gamma$, $\alpha$, $\beta$, $\gamma$ described above;
6 parameters ($\Re_{\eta S}$, $\Im_{\eta S}$, $\Re_{\eta P}$, $\Im_{\eta P}$, 
$\Re_{\eta D}$, $\Im_{\eta D}$) characterizing the interference between the signal
  amplitude $A_{\eta}$ and background amplitudes $A_{S,P,D}$;
3 parameters ($\Pi_{SP}$, $\Pi_{SD}$, $\Pi_{PD}$) describing the contributions 
from the interference between the background amplitudes $A_{S,P,D}$. 
In particular,
\begin{equation}
\Re_{\eta i}+i\Im_{\eta i}=
\int\int\int A_{\eta}(q_1^2,q_2^2)A_i^*(q_1^2,q_2^2)dq_1^2dq_2^2d\phi,
\end{equation}
\begin{equation}
\Pi_{ij}=
\int\int\int\Re\left(A_i(q_1^2,q_2^2)A_j^*(q_1^2,q_2^2)\right)dq_1^2dq_2^2d\phi,
\end{equation}
where $i,j=S,P,D$ and $i\neq j$.

Since the function $F(s,x)$ is a sum of the $\eta_c$ 
($\eta_c(2S)$) 
Breit-Wigner and S-, P-, and D-waves, it can be represented as a rational 
function of $s$ and $x$:
\begin{equation}
F(s,x)=\frac{1+\varepsilon_1 x+\varepsilon_2 x^2}{(s-M^2)^2+M^2\Gamma^2}\sum_{i=0}^{2}\sum_{j=0}^{4}C_{ij}s^ix^j.
\end{equation}
In its most general form, such a function has 15 independent
coefficients ($C_{ij}$) in the numerator and two ($M$ and $\Gamma$) in the 
denominator, however, in our case some coefficients are not independent:

\begin{minipage}[t]{73mm}
1. $C_{03}=M^2(M^2+\Gamma^2)C_{23}$, 

2. $C_{13}=-2M^2C_{23}$,
\end{minipage}
\begin{minipage}[t]{73mm}
3. $C_{04}=M^2(M^2+\Gamma^2)C_{24}$, 

4. $C_{14}=-2M^2C_{24}$.

\end{minipage}

Thus, we have ($15+2-4=13$) independent terms only, which is not
enough to determine  all 15 parameters of the function $F(s,x)$. 
Two ($15-13=2$) of
the parameters must be either obtained from other measurements
or allowed to vary over the full allowed range.
Since we have no additional information on these parameters, 
we scan over them. The result does not depend on the choice of the two scanned
parameters. Since the interference is more significant in the S-wave, we choose
$\alpha$ and $\Im_{\eta S}$. To perform this scan, we randomly
sample $\alpha$ and $\Im_{\eta S}$ in a reasonable range\footnote{The scan range
of parameter $\alpha$ is chosen so that it includes the minimum $\chi^2$ 
region. It is [0:9] for the $\eta_c$ and [1:11] for the $\eta_c(2S)$ analyses.
The parameter $\Im_{\eta S}$ is varied over the entire physical region [-1:1].} with
the remaining 13 parameters free. After fitting the distributions we obtain 
a set of parameters and a $\chi^2$. No additional local minima are found.

\begin{figure}[!ht]
\begin{center}
\begin{tabular}{cc}
\includegraphics[height=5 cm]{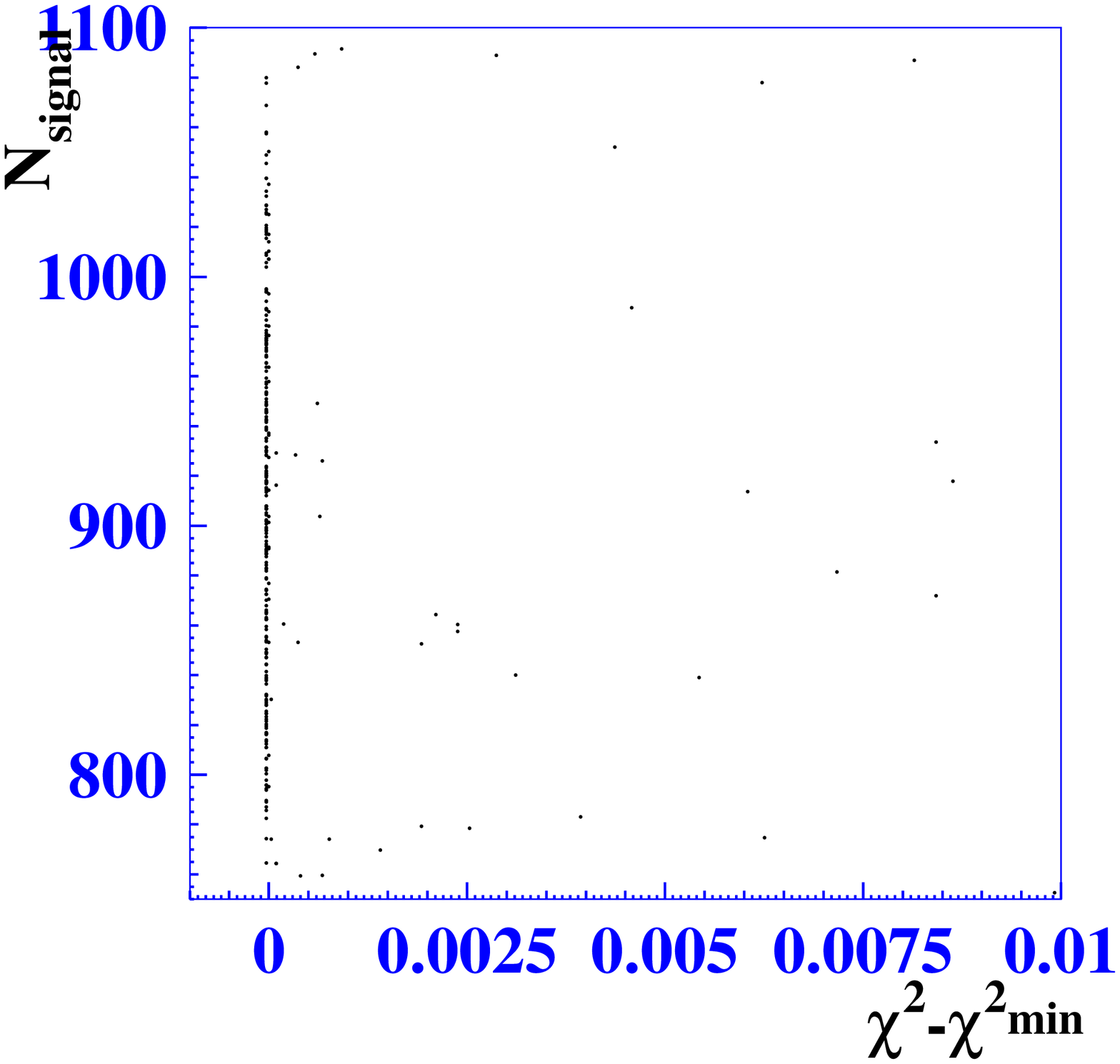}&
\includegraphics[height=5 cm]{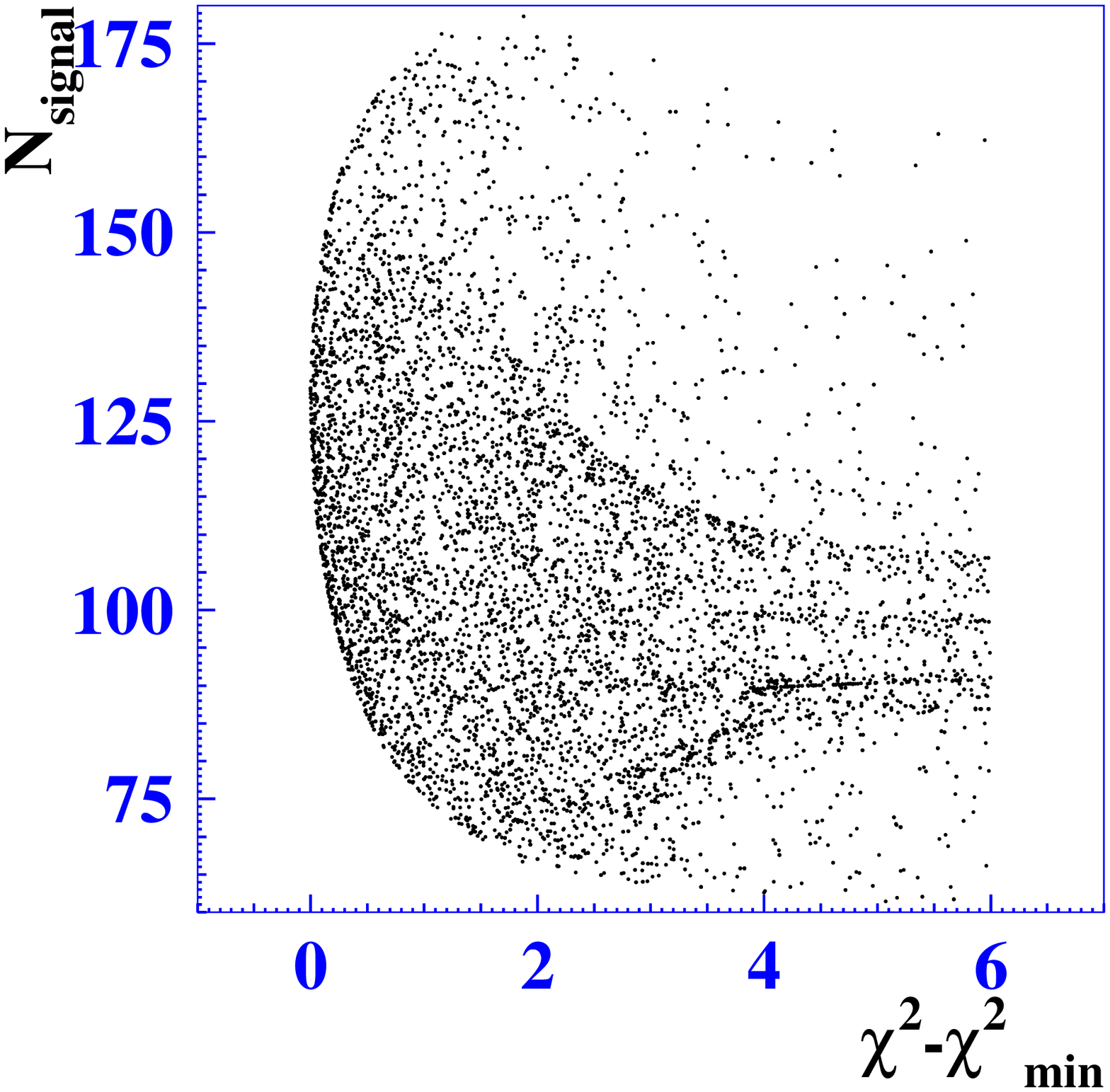}\\
\end{tabular}
\caption{The dependences of the number of signal events on $\chi^2$ for 
the $\eta_c$ (left) and $\eta_c(2S)$ (right) decay analyses. One should note
that the $\chi^2$ axis of the left plot covers a much smaller range than
that of the right plot.}
\label{pic:chi}
\end{center}
\end{figure}

The dependences of the signal yields 
on $\chi^2$ are shown in Fig.~\ref{pic:chi}.
In the $\eta_c$ case,
one can see that this distribution has a "plateau", which
consists of fits with different $N_{signal}$ (and other fit parameters) 
and the same $\chi^2$. This feature arises because
our system of equations for the fit parameters is underdetermined.  
The variation of 
parameters within this plateau will be referred to as the model uncertainty of 
our analysis and
the average of their statistical errors as the statistical uncertainty.
In the $\eta_c(2S)$ case, the minimum $\chi^2$ plateau is not reached, because 
the parameters $\Re_{\eta S,\eta P,\eta D}$, $\Im_{\eta S,\eta P,\eta D}$, and 
$\Pi_{SP,SD,PD}$ tend to 
their bounds\footnote{By definition, $\Re_{\eta S,\eta P,\eta D}$, 
$\Im_{\eta S,\eta P,\eta D}$, and
$\Pi_{SP,SD,PD}$ vary in the interval 
[-1:1].}. These bounds make our system of equations fully determined, so
we can float the parameters $\alpha$ and $\Im_{\eta S}$.
In that case, the model and statistical errors cannot be separated.
Thus we obtain
$N_{signal}=920\pm 50(stat)\pm 170(model)$ for the $\eta_c$ decay and
$N_{signal}=128^{+83}_{-58}(stat+model)$ for the $\eta_c(2S)$ decay.

\begin{figure}[!h]
\begin{center}
\begin{tabular}{ccc}
\includegraphics[height=4 cm]{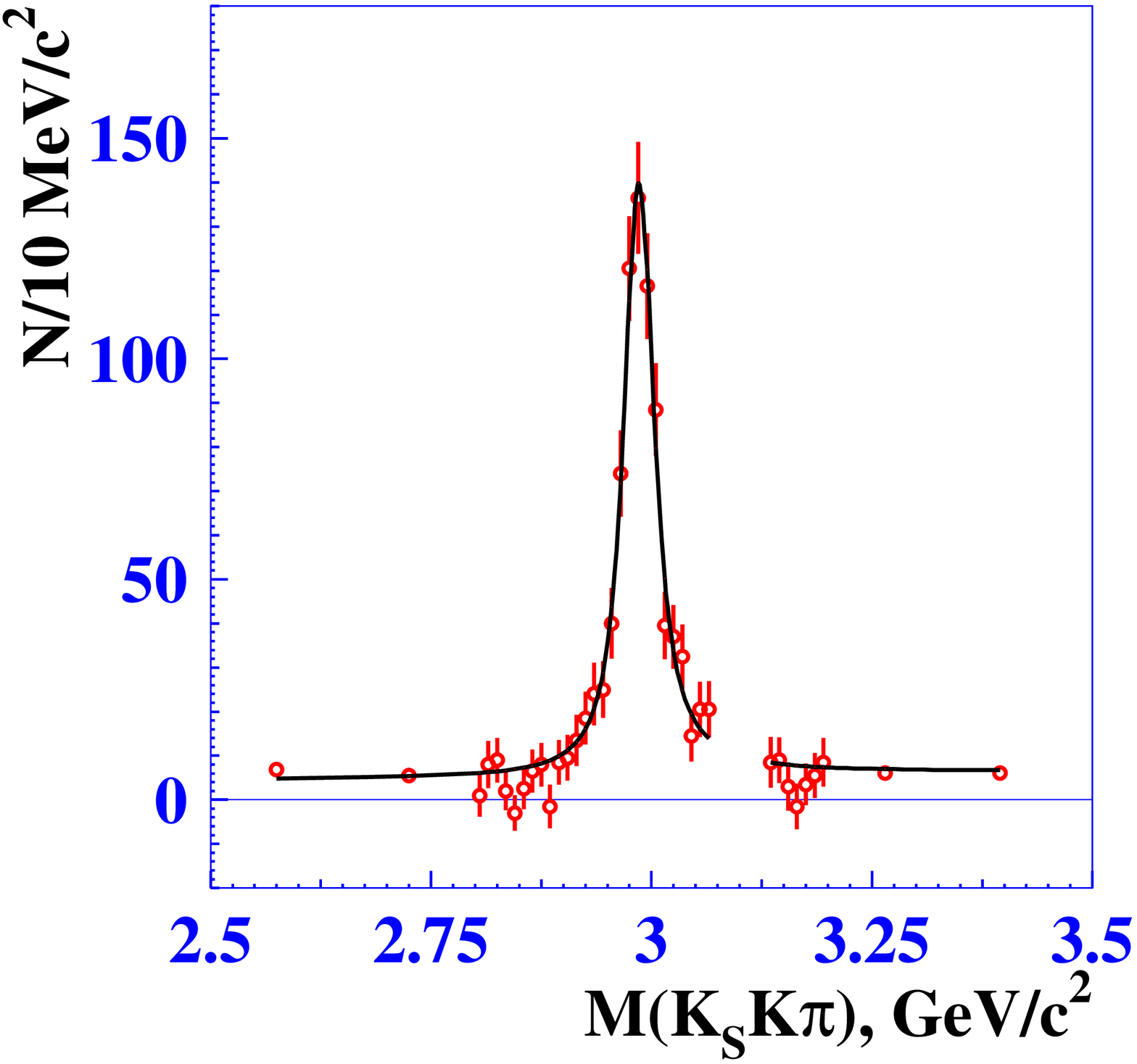}&
\includegraphics[height=4 cm]{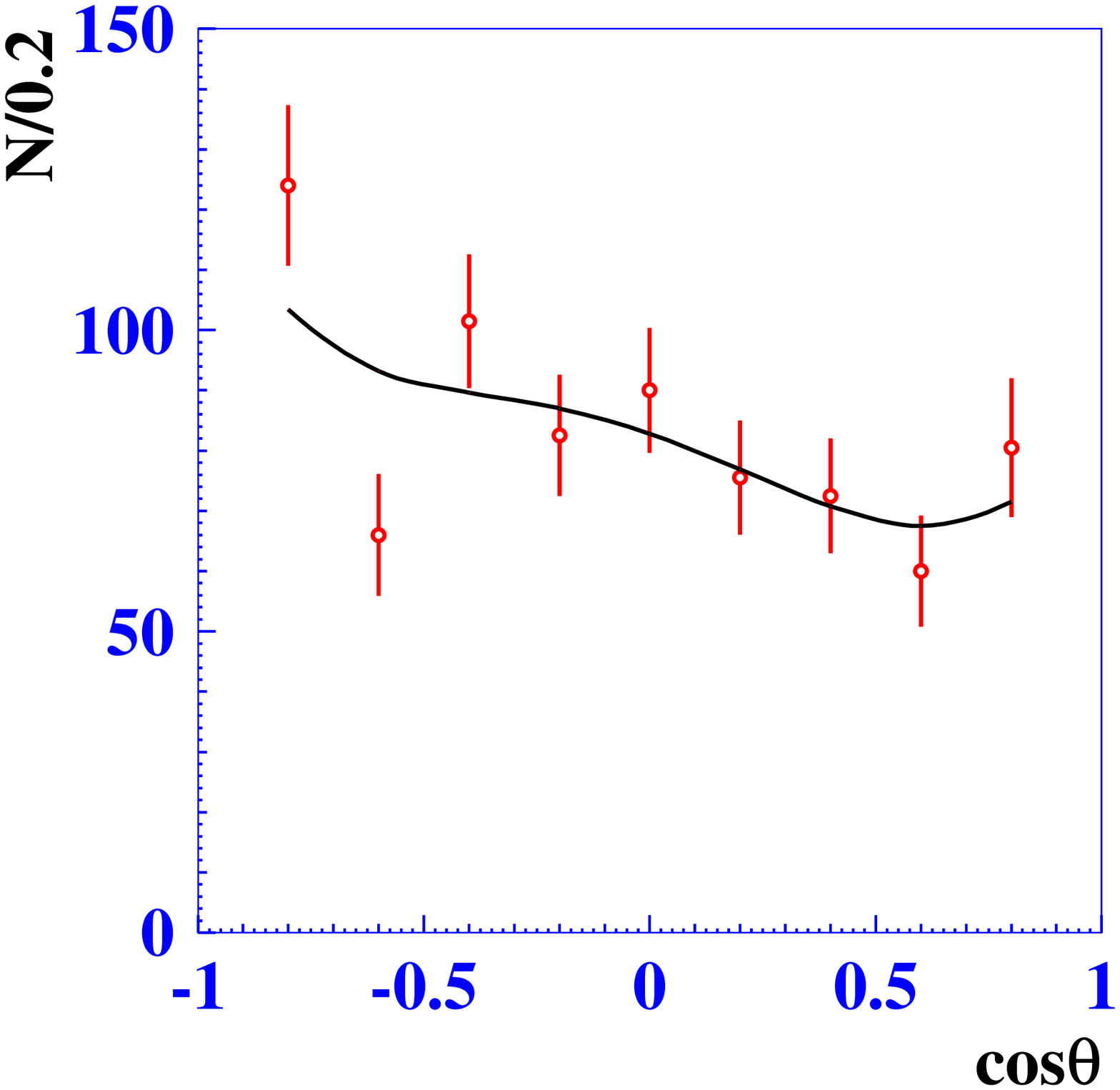}&
\includegraphics[height=4 cm]{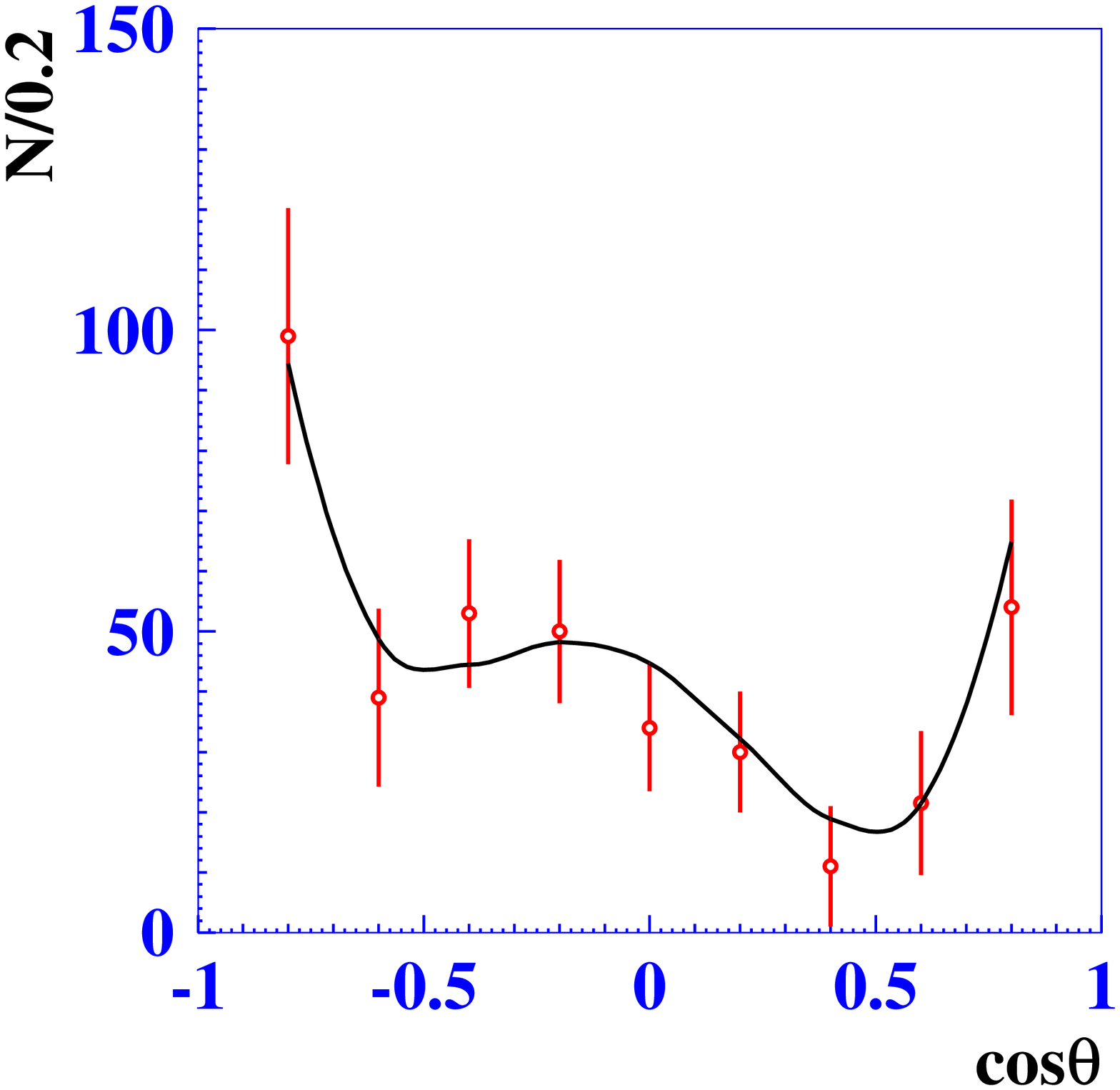}\\
\end{tabular}
\caption{Projections of the fit in $K_SK\pi$
invariant mass in the $\eta_c$ mass region (left) and $\cos\theta$ in the $\eta_c$ invariant 
mass signal (center) and sideband (right) regions.
The combinatorial background
is subtracted. The gap near 3.1 GeV/$c^2$ is due to the $J/\psi$ veto.
The bin size along the
$\cos{\theta}$ axis is $0.2$. Along the $M(K_S K\pi)$ axis the bin size is
$10$ MeV/$c^2$ in
the signal region and $150/130$ MeV/$c^2$ in the left/right sideband 
region}
\label{pic:fit1}
\end{center}
\end{figure}

\begin{figure}[!h]
\begin{center}
\begin{tabular}{ccc}
\includegraphics[height=4 cm]{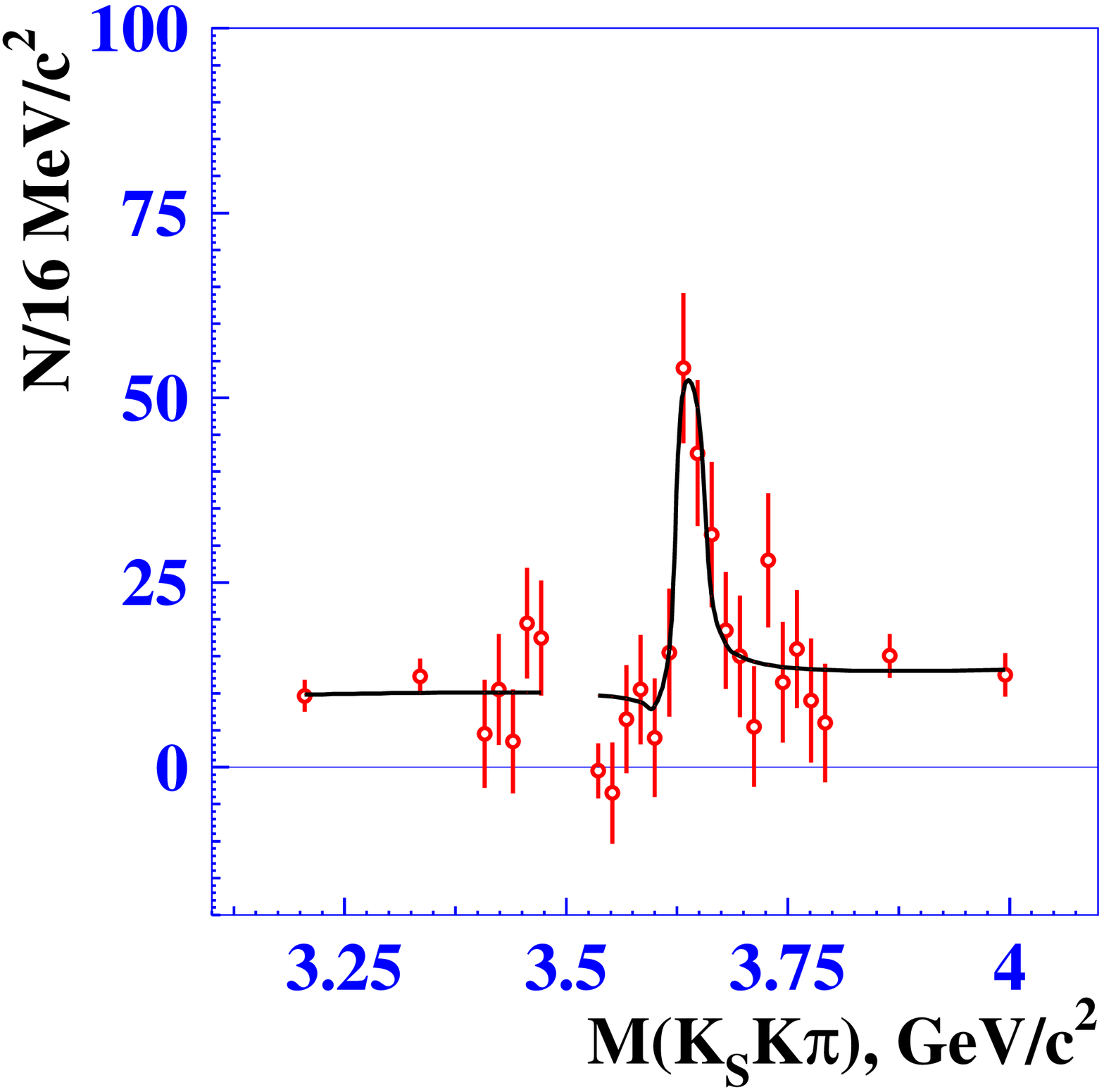}&
\includegraphics[height=4 cm]{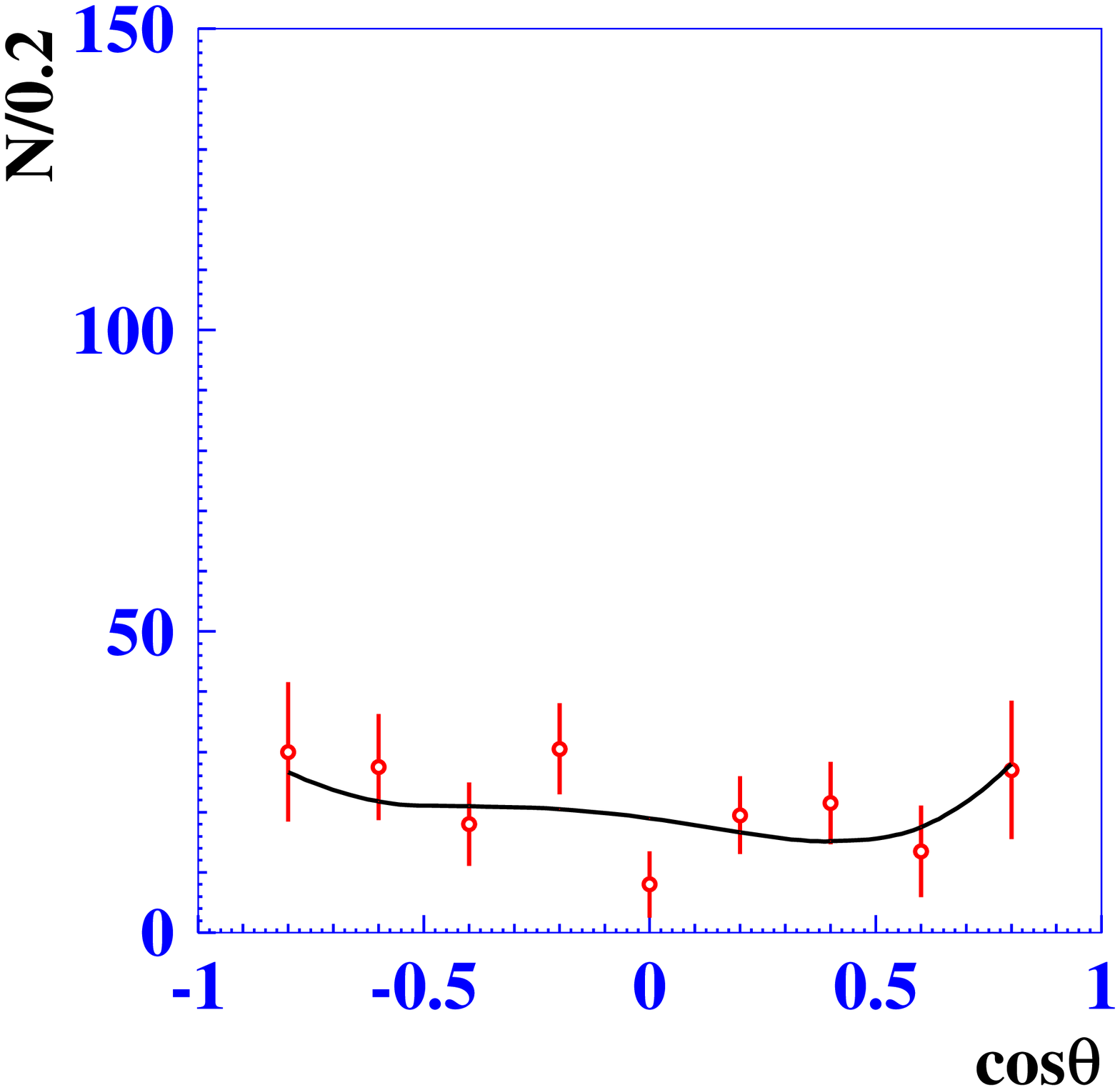}&
\includegraphics[height=4 cm]{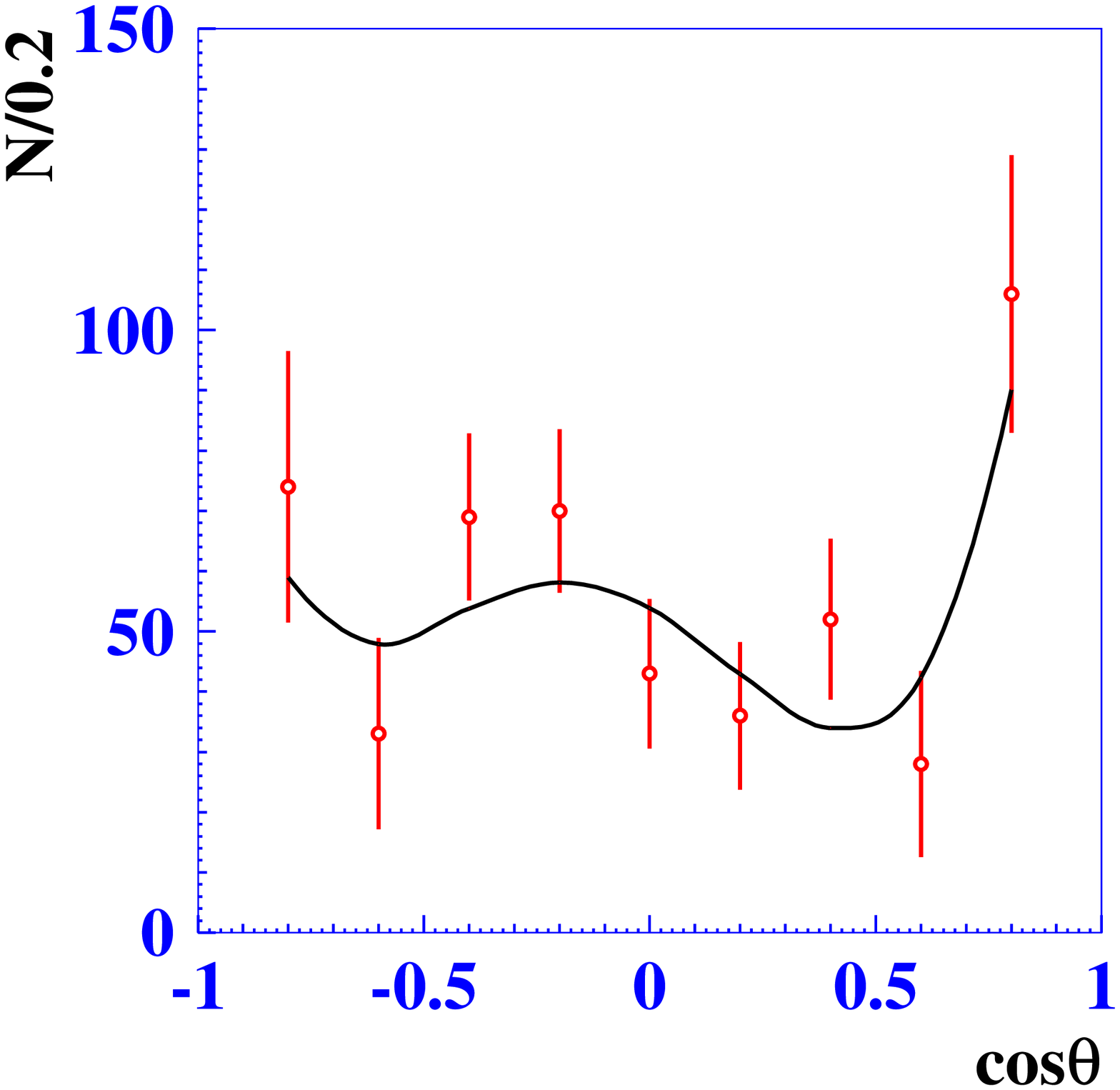}\\
\end{tabular}
\caption{Projections of the fit in $K_SK\pi$ invariant mass in the 
$\eta_c(2S)$ mass region (left) and
$\cos\theta$ in the $\eta_c(2S)$ invariant 
mass signal (center) and sideband (right) regions. 
The combinatorial background
is subtracted. The gap near 3.5 GeV/$c^2$ is due to the $\chi_{c1}$ veto.
The bin size along the
$\cos{\theta}$ axis is $0.2$. Along the $M(K_S K\pi)$ axis the bin size is 
$16$ MeV/$c^2$ in the signal region and $130$ MeV/$c^2$ in the sideband region.}
\label{pic:fit2}
\end{center}
\end{figure}

Projections of the fits using the function $F(s,x)$ are shown in
Figs.~\ref{pic:fit1} and~\ref{pic:fit2}.

The fit procedure described above was also applied to MC signal samples.
The obtained number of signal events was used to determine the detection
efficiency ($(9.32\pm 0.10)$\% and $(10.18\pm0.10)$\% for $\eta_c$ and 
$\eta_c(2S)$ decays, respectively) and hence to calculate the product branching 
fractions.

The decay $B^{\pm}\to K^{\pm}(K_SK\pi)^0$ has
two possible final states: $K^{\pm}K_SK^{\pm}\pi^{\mp}$ and 
$K^{\pm}K_SK^{\mp}\pi^{\pm}$. We
assume that the $\eta_c$ decay signal amplitudes are the same, but the 
non-resonant contributions ($\alpha$, $\beta$, $\gamma$) could be different in 
each decay channel. Because of limited statistics we do not treat these
final states separately.
A single distribution~(\ref{eq:f}) effectively describes the 
incoherent sum of two distributions for the two decay channels. This does 
not affect the parameters of $\eta_c$ ($\eta_c(2S)$) states, but the parameters 
describing the non-resonant part obtained in the fit take 
effective values that depend on both non-resonant amplitudes.

The method described above was checked using toy MC, which showed that the 
described procedure gives parameter values consistent with the 
generated ones. 
Moreover, a generic MC test was performed that included a full
simulation of all $b\to c$ decays at the $\Upsilon(4S)$ without interference.
This test verified that the signal determination is not biased
and gave an interference value consistent with zero.

\section{Systematic uncertainties}

We evaluated possible sources of systematic uncertainties in the product
branching fractions.
The number of $B\bar{B}$ pairs is calculated from the difference of the 
number of hadronic events on resonance and the scaled number of those 
off-resonance. The systematic error is dominated by the uncertainty in the scale 
factor and is equal to $\sim$1.3\%.
We assume that the combinatorial background can be parameterized with a 
first-order polynomial. To obtain the background shape uncertainty, 
we describe the background by a second-order polynomial and compare the results. 
The uncertainty on the $K_S$ decay branching fraction is taken from~\cite{PDG}.
The contribution of the $K_S$ reconstruction uncertainty was estimated in the
Belle experiment to be 4.4\%~\cite{Ks}.
In our fitting procedure we take into account the efficiency dependence on 
$\cos\theta$ and assume that it does not depend on
the $K_SK\pi$ invariant mass. By adding a linear dependence on $M(K_SK\pi)$
we estimate the corresponding systematic error. Moreover, we take into account
the dependence of the efficiency on the $\eta_c$ and $\eta_c(2S)$ decay models, such 
as $KK^*$,
$KK^*_0(1430)$, and $KK^*_2(1430)$. The corresponding contribution to the 
systematic uncertainty is estimated
by varying the efficiency obtained using these models and taking the
difference in the results.  
An analysis of the charged track reconstruction
uncertainty as a function of particle momenta has been performed in 
Belle data and gave an estimate of 1\% per charged track. 
To determine the errors due to $K$ and $\pi$ meson identification, data from
analysis of the process $D^{*+}\to D^0\pi^+$ followed by the decay
$D^0\to K^-\pi^+$ were used. The
uncertainty in $K^{\pm}$ identification is 0.8\% per $K$ meson and 
the corresponding value for $\pi^{\pm}$ identification is 0.5\% per $\pi$ meson.  
We also take into account the deviation of MC from the data by applying a correction
to the efficiency:
$\frac{\varepsilon_{Data}}{\varepsilon_{MC}}$ is 0.9996 for each kaon
and 0.9756 for each pion.
We vary the bin size along the
x axis from 0.15 to 0.225, the $\Delta E$ window from 20 MeV to 40 MeV, and the detector resolution
 within the limits of its statistical uncertainty.

Sources of systematic uncertainties for masses and widths include
the background parameterization, bin size, and detector resolution as
described above, as well as a scale uncertainty and the effect of identical
kaons in the final state. The
scale uncertainty is determined from a comparison of masses of the $J/\psi$ ($\chi_{c1}$)
resonances, which were obtained by fitting the $K_SK\pi$ invariant mass 
distribution, with the world average values~\cite{PDG}.
In case two kaons of the same charge are present in the final state, the 
charmonium amplitude is a sum of two amplitudes corresponding to
$K_SK_{(1)}\pi$ and $K_SK_{(2)}\pi$ combinations. This can lead to the
deformation of the $K_SK\pi$ invariant mass distribution in the region of
phase space where the $K_SK_{(1)}\pi$ and $K_SK_{(2)}\pi$ invariant mass values
overlap. We use toy MC to estimate this effect and take it
into account as an additional systematic error.

All the contributions to the systematic uncertainties are listed in
Table~\ref{tab:syst1} for the product branching fractions and in Table~\ref{tab:syst2}
for the masses and widths.
\begin{table}[!h]
\begin{center}
\caption{Systematic uncertainties of the product branching fractions (in \%).}
\begin{tabular}{|c||c|c|} 
\hline
Source &
\multicolumn{2}{c|}{$B^{\pm}\to
  K^{\pm}(K_S K\pi)^0$}\\ 
\cline{2-3}
 & ~~~~$\eta_c$~~~~ &
$\eta_c(2S)$\\
\hline \hline
Number of $B\bar{B}$ pairs & $1.3$ & $1.3$\\ 
${\mathcal B}(K_S\to\pi^+\pi^-)$ & $0.1$ & $0.1$\\
Model efficiency dependence & $^{+8.6}_{-6.7}$ & $^{+2.0}_{-1.5}$\\
Background approximation & --- & $+2.3$\\
Bin size & $-3.3$ & $^{+13.3}_{-3.9}$\\
$\Delta E$ cut & $-2.2$ & $+2.3$ \\
Detector resolution & $+1.1$ & $^{+4.7}_{-8.6}$\\ 
$M_{inv}$ efficiency dependence & $+2.2$ & $+0.8$ \\
Track reconstruction & $3$ & $3$\\
$K^{\pm}$ identification & $1.6$ & $1.6$\\
$\pi^{\pm}$ identification & $1.5$ & $1.5$\\
$K_S$ reconstruction & $4.4$ & $4.4$\\ 
\hline \hline
Total, \% & $^{+10.7}_{-9.8}$ & $^{+15.8}_{-11.9}$\\
\hline
\end{tabular}
\label{tab:syst1} 
\end{center}
\end{table}
\begin{table}[!h]
\begin{center}
\caption{Systematic uncertainties of masses and widths of
  the $\eta_c$ and $\eta_c(2S)$ mesons (in MeV/$c^2$).}
\begin{tabular}{|c||c|c||c|c|} 
\hline
Source & \multicolumn{2}{c||}{$\eta_c$}
& \multicolumn{2}{|c|}{$\eta_c(2S)$}\\
\cline{2-5}
 & ~~Mass~~ &
~Width~ & ~~Mass~~ & ~Width~~\\
\hline\hline
Background approximation & --- & --- & $+0.2$ &
$-0.1$\\
Bin size & $+0.2$ & $-1.0$ & $-1.1$ & $+2.4$\\
Detector resolution & $-0.1$ & $^{+1.0}_{-1.2}$ &
$^{+0.5}_{-0.1}$ & $^{+1.0}_{-0.9}$\\
Scale uncertainty & $-2.0$ & --- & $-1.7$ & ---\\
Effect of identical kaons & $+0.5$ & $+0.3$ & $+0.5$ & $+0.3$\\
\hline\hline
Total, MeV/$c^2$ & $^{+0.5}_{-2.0}$ & $^{+1.0}_{-1.6}$ & $^{+0.7}_{-2.0}$ &
$^{+2.6}_{-0.9}$\\ 
\hline
\end{tabular}
\label{tab:syst2}
\end{center}
\end{table}

\section{Results and discussion}

Table~\ref{tab:res} shows a comparison of the results obtained assuming
no interference (1-D fits to the $\Delta E$ and to the $K_SK\pi$ invariant 
mass distributions) and those obtained using
the analysis described above. One can see that taking interference
into account leads to
the introduction of a model error for the product branching fractions 
${\mathcal B}(B^{\pm}\to K^{\pm}\eta_c){\mathcal B}(\eta_c\to K_S K^{\pm}\pi^{\mp})$ 
(for the $\eta_c$ mass and width this error turns out to be negligibly
small). In the $\eta_c(2S)$ decay analysis the model error is not listed
separately, but the results differ noticeably from those that assume no 
interference.
\begin{table}[!h]
\begin{center}
\caption{Comparison of the results obtained under the assumption of
  no interference between the signal and the non-resonant contribution
  and those obtained with interference.}
\begin{tabular}{|c||c|c|} 
\hline
 & \footnotesize{No interference} & \footnotesize{Taking interference into account}\\
\hline\hline
\multicolumn{3}{|c|}{\footnotesize{$B^{\pm}\to K^{\pm}\eta_c$,
$\eta_c\to (K_S K\pi)^0$}}\\
\hline
\footnotesize{${\mathcal B}\times {\mathcal B}$, $10^{-6}$} & \footnotesize{$24.0\pm
1.2(stat)^{+2.1}_{-2.0}(syst)$} & \footnotesize{$26.7\pm
1.4(stat)^{+2.9}_{-2.6}(syst)\pm 4.9(model)$}\\
\footnotesize{Mass, MeV/$c^2$} & \footnotesize{$2984.8\pm 1.0(stat)^{+0.1}_{-2.0}(syst)$} & \footnotesize{$2985.4\pm
1.5(stat)^{+0.5}_{-2.0}(syst)$}\\
\footnotesize{Width, MeV/$c^2$} & \footnotesize{$35.4\pm 3.6(stat)^{+3.0}_{-2.1}(syst)$} & \footnotesize{$35.1\pm
3.1(stat)^{+1.0}_{-1.6}(syst)$}\\
\hline\hline
\multicolumn{3}{|c|}{\footnotesize{$B^{\pm}\to K^{\pm}\eta_c(2S)$,
$\eta_c(2S)\to (K_S K\pi)^0$}}\\
\hline
\footnotesize{${\mathcal B}\times {\mathcal B}$, $10^{-6}$} & \footnotesize{$3.1\pm
0.8(stat)\pm 0.2(syst)$} & \footnotesize{$3.4
^{+2.2}_{-1.5}(stat+model)^{+0.5}_{-0.4}(syst)$}\\
\footnotesize{Mass, MeV/$c^2$} & \footnotesize{$3646.5\pm 3.7(stat)^{+1.2}_{-2.9}(syst)$} & \footnotesize{$3636.1
 ^{+3.9}_{-4.2}(stat+model)^{+0.7}_{-2.0}(syst)$}\\
\footnotesize{Width, MeV/$c^2$} & \footnotesize{$41.1\pm 12.0 (stat)^{+6.4}_{-10.9}(syst)$} &
 \footnotesize{$6.6^{+8.4}_{-5.1}(stat+model)^{+2.6}_{-0.9}(syst)$}\\
\hline
\end{tabular}
\label{tab:res}
\end{center}
\end{table}

Table~\ref{tab:comp2} shows that there is a large spread in the 
$\eta_c(2S)$ width values. A possible explanation of this spread is that
the previous studies did not take interference into account. For each
of the studied processes the interference could have a different effect on the
results and shift the $\eta_c(2S)$ mass value significantly.
Thus it is important to take interference into account.

In addition to affecting the value of the branching fraction, interference
changes the Breit-Wigner shape. This effect can allow the improvement of the
statistical accuracy with which the Breit-Wigner width is determined. In 
particular,
the $\eta_c(2S)$ width, obtained in the present work, has a rather good 
accuracy, despite limited statistics and a detector resolution
broader than the intrinsic width. The
interference deforms the Breit-Wigner, lengthening its tail and thus
improves the fit to the width (see Fig.~\ref{pic:fit2}). 

\section{Conclusion}

We report a study of the decay $B^{\pm}\to K^{\pm}(c\bar{c})$, where the 
$(c\bar{c})$ state decays to $(K_SK\pi)^0$ and includes the $\eta_c$
and $\eta_c(2S)$ charmonia states. Both decay channels contain 
$B^{\pm}\to K^{\pm}(K_SK\pi)^0$ decays without intermediate charmonia
that interfere with the signal. For the first time, the analysis takes 
interference into account with no assumptions on
the phase or absolute value of the interference.

As a result, we obtain an estimate of the model error for 
${\mathcal B}(B^{\pm}\to K^{\pm}\eta_c){\mathcal B}(\eta_c\to K_S K^{\pm}\pi^{\mp})$:
${\mathcal B}(B^{\pm}\to K^{\pm}\eta_c){\mathcal B}(\eta_c\to K_S K^{\pm}\pi^{\mp})
=(26.7\pm 1.4(stat)^{+2.9}_{-2.6}(syst)\pm 4.9(model))\times 10^{-6}$.
For ${\mathcal B}(B^{\pm}\to K^{\pm}\eta_c(2S)){\mathcal B}(\eta_c(2S)\to 
K_S K^{\pm}\pi^{\mp})$, the model error from interference is not 
listed separately:
${\mathcal B}(B^{\pm}\to K^{\pm}\eta_c(2S)){\mathcal B}(\eta_c(2S)\to 
K_S K^{\pm}\pi^{\mp})=(3.4^{+2.2}_{-1.5}(stat+model)^{+0.5}_{-0.4}(syst))
\times 10^{-6}$.

We also obtain the masses and widths of $\eta_c$ and $\eta_c(2S)$.
For the $\eta_c$ meson parameters the model error is negligibly small:\\
$M(\eta_c)=2985.4\pm1.5(stat)^{+0.5}_{-2.0}(syst)$ MeV/$c^2$,\\
$\Gamma(\eta_c)=35.1\pm 3.1(stat)^{+1.0}_{-1.6}(syst)$ MeV/$c^2$.\\
For the $\eta_c(2S)$ meson the model and statistical uncertainties cannot
be separated:\\
$M(\eta_c(2S))=3636.1^{+3.9}_{-4.2}(stat+model)^{+0.7}_{-2.0}(syst)$ MeV/$c^2$,\\
$\Gamma(\eta_c(2S))=6.6^{+8.4}_{-5.1}(stat+model)^{+2.6}_{-0.9}(syst)$ 
MeV/$c^2$.\\
For the $\eta_c(2S)$ the interference has a dramatic effect on the
width (see Table~\ref{tab:res}).

These results are consistent with those obtained in the most accurate
existing measurements. Our errors are comparable to those in other experiments
despite the fact that they include additional uncertainty related to
interference effects.

\section*{Acknowledgments}

We thank the KEKB group for the excellent operation of the
accelerator, the KEK cryogenics group for the efficient
operation of the solenoid, and the KEK computer group and
the National Institute of Informatics for valuable computing
and SINET3 network support.  We acknowledge support from
the Ministry of Education, Culture, Sports, Science, and
Technology (MEXT) of Japan, the Japan Society for the 
Promotion of Science (JSPS), and the Tau-Lepton Physics 
Research Center of Nagoya University; 
the Australian Research Council and the Australian 
Department of Industry, Innovation, Science and Research;
the National Natural Science Foundation of China under
contract No.~10575109, 10775142, 10875115 and 10825524; 
the Ministry of Education, Youth and Sports of the Czech 
Republic under contract No.~LA10033 and MSM0021620859;
the Department of Science and Technology of India; 
the BK21 and WCU program of the Ministry Education Science and
Technology, National Research Foundation of Korea,
and NSDC of the Korea Institute of Science and Technology Information;
the Polish Ministry of Science and Higher Education;
the Ministry of Education and Science of the Russian
Federation and the Russian Federal Agency for Atomic Energy;
the Slovenian Research Agency;  the Swiss
National Science Foundation; the National Science Council
and the Ministry of Education of Taiwan; and the U.S.\
Department of Energy.
This work is supported by a Grant-in-Aid from MEXT for 
Science Research in a Priority Area (``New Development of 
Flavor Physics''), and from JSPS for Creative Scientific 
Research (``Evolution of Tau-lepton Physics'').

\section*{Appendix}

After raising the absolute value to the second power $F(s,x)$ 
(Eq.~\ref{eq:f}) convolved with the resolution Gaussian function can be
written as: 
\begin{eqnarray}
F(s,x)&=&\int_{x-\frac{\delta}{2}}^{x+\frac{\delta}{2}}(1+\varepsilon_1
x'+\varepsilon_2 x'^2)
\Biggl[S^2(x')\left(I_{\eta\eta}(s)+\alpha^2\Delta+2\sqrt{N}\alpha I_{\eta
    S}(s)\right)+\nonumber\\
& &P^2(x')\beta^2\Delta+D^2(x')\gamma^2\Delta+2S(x')P(x')\beta\left(\sqrt{N}I_{\eta
    P}(s)+\alpha\Pi_{SP}\Delta\right)+ \nonumber\\
& &2S(x')D(x')\gamma\left(\sqrt{N}I_{\eta_D}(s)+\alpha\Pi_{SD}\Delta\right)+2P(x')D(x')\beta\gamma\Pi_{PD}\Delta\Biggr]dx',
\end{eqnarray}
where we use the following notations:
\begin{eqnarray}
I_{\eta\eta}(s)&=&\int\int\int\int_{s-\frac{\Delta}{2}}^{s+\frac{\Delta}{2}}
\left|\frac{\sqrt{N}}{s'-M^2+iM\Gamma}A_{\eta}(q_1^2,q_2^2)\right|^2\otimes\frac{\exp(-\frac{s'^2}{2\sigma^2})}{\sqrt{2\pi}\sigma}ds'dq_1^2dq_2^2d\phi=
\nonumber\\
& &\int_{s-\frac{\Delta}{2}}^{s+\frac{\Delta}{2}}
\left|\frac{\sqrt{N}}{s'-M^2+iM\Gamma}\right|^2\otimes\frac{\exp(-\frac{s'^2}{2\sigma^2})}{\sqrt{2\pi}\sigma}ds',
\end{eqnarray}
\begin{eqnarray}
I_{\eta S}(s)&=&\int\int\int\int_{s-\frac{\Delta}{2}}^{s+\frac{\Delta}{2}}
\Re\left(\frac{1}{s'-M^2+iM\Gamma}A_{\eta}(q_1^2,q_2^2)A_S^*(q_1^2,q_2^2)\right)\otimes\nonumber\\
& &\frac{\exp(-\frac{s'^2}{2\sigma^2})}{\sqrt{2\pi}\sigma}ds'dq_1^2dq_2^2d\phi=
\nonumber\\
& &\Re_{\eta
  S}\Re\left[\int_{s-\frac{\Delta}{2}}^{s+\frac{\Delta}{2}}\left(\frac{1}{s'-M^2+iM\Gamma}\right)\otimes\frac{\exp(-\frac{s'^2}{2\sigma^2})}{\sqrt{2\pi}\sigma}ds'\right]+ \nonumber\\
& &\Im_{\eta
  S}\Im\left[\int_{s-\frac{\Delta}{2}}^{s+\frac{\Delta}{2}}\left(\frac{1}{s'-M^2+iM\Gamma}\right)\otimes\frac{\exp(-\frac{s'^2}{2\sigma^2})}{\sqrt{2\pi}\sigma}ds'\right],
\end{eqnarray}
\begin{eqnarray}
I_{\eta P}(s)&=&\int\int\int\int_{s-\frac{\Delta}{2}}^{s+\frac{\Delta}{2}}
\Re\left(\frac{1}{s'-M^2+iM\Gamma}A_{\eta}(q_1^2,q_2^2)A_P^*(q_1^2,q_2^2)\right)\otimes\nonumber\\
& &\frac{\exp(-\frac{s'^2}{2\sigma^2})}{\sqrt{2\pi}\sigma}ds'dq_1^2dq_2^2d\phi=
\nonumber\\
& &\Re_{\eta
  P}\Re\left[\int_{s-\frac{\Delta}{2}}^{s+\frac{\Delta}{2}}\left(\frac{1}{s'-M^2+iM\Gamma}\right)\otimes\frac{\exp(-\frac{s'^2}{2\sigma^2})}{\sqrt{2\pi}\sigma}ds'\right]+ \nonumber\\
& &\Im_{\eta
  P}\Im\left[\int_{s-\frac{\Delta}{2}}^{s+\frac{\Delta}{2}}\left(\frac{1}{s'-M^2+iM\Gamma}\right)\otimes\frac{\exp(-\frac{s'^2}{2\sigma^2})}{\sqrt{2\pi}\sigma}ds'\right],
\end{eqnarray}
\begin{eqnarray}
I_{\eta D}(s)&=&\int\int\int\int_{s-\frac{\Delta}{2}}^{s+\frac{\Delta}{2}}
\Re\left(\frac{1}{s'-M^2+iM\Gamma}A_{\eta}(q_1^2,q_2^2)A_D^*(q_1^2,q_2^2)\right)\otimes\nonumber\\
& &\frac{\exp(-\frac{s'^2}{2\sigma^2})}{\sqrt{2\pi}\sigma}ds'dq_1^2dq_2^2d\phi=
\nonumber\\
& &\Re_{\eta
  D}\Re\left[\int_{s-\frac{\Delta}{2}}^{s+\frac{\Delta}{2}}\left(\frac{1}{s'-M^2+iM\Gamma}\right)\otimes\frac{\exp(-\frac{s'^2}{2\sigma^2})}{\sqrt{2\pi}\sigma}ds'\right]+ \nonumber\\
& &\Im_{\eta
  D}\Im\left[\int_{s-\frac{\Delta}{2}}^{s+\frac{\Delta}{2}}\left(\frac{1}{s'-M^2+iM\Gamma}\right)\otimes\frac{\exp(-\frac{s'^2}{2\sigma^2})}{\sqrt{2\pi}\sigma}ds'\right],
\end{eqnarray}
\begin{equation}
\Re_{\eta S}=
\int\int\int\Re\left(A_{\eta}(q_1^2,q_2^2)A_S^*(q_1^2,q_2^2)\right)dq_1^2dq_2^2d\phi=\xi_{\eta S}\sqrt{1-\Im_{\eta S}^2},
\end{equation}
\begin{equation}
\Im_{\eta S}=
\int\int\int\Im\left(A_{\eta}(q_1^2,q_2^2)A_S^*(q_1^2,q_2^2)\right)dq_1^2dq_2^2d\phi,
\end{equation}
\begin{equation}
\Re_{\eta P}=
\int\int\int\Re\left(A_{\eta}(q_1^2,q_2^2)A_P^*(q_1^2,q_2^2)\right)dq_1^2dq_2^2d\phi=\xi_{\eta P}\cos\theta_{\eta P},
\end{equation}
\begin{equation}
\Im_{\eta P}=
\int\int\int\Im\left(A_{\eta}(q_1^2,q_2^2)A_P^*(q_1^2,q_2^2)\right)dq_1^2dq_2^2d\phi=\xi_{\eta P}\sin\theta_{\eta P},
\end{equation}
\begin{equation}
\Re_{\eta D}=
\int\int\int\Re\left(A_{\eta}(q_1^2,q_2^2)A_D^*(q_1^2,q_2^2)\right)dq_1^2dq_2^2d\phi=\xi_{\eta D}\cos\theta_{\eta D},
\end{equation}
\begin{equation}
\Im_{\eta D}=
\int\int\int\Im\left(A_{\eta}(q_1^2,q_2^2)A_D^*(q_1^2,q_2^2)\right)dq_1^2dq_2^2d\phi=\xi_{\eta D}\sin\theta_{\eta D},
\end{equation}
\begin{equation}
\Pi_{SP}=
\int\int\int\Re\left(A_S(q_1^2,q_2^2)A_P^*(q_1^2,q_2^2)\right)dq_1^2dq_2^2d\phi,
\end{equation}
\begin{equation}
\Pi_{SD}=
\int\int\int\Re\left(A_S(q_1^2,q_2^2)A_D^*(q_1^2,q_2^2)\right)dq_1^2dq_2^2d\phi,
\end{equation}
\begin{equation}
\Pi_{PD}=
\int\int\int\Re\left(A_P(q_1^2,q_2^2)A_D^*(q_1^2,q_2^2)\right)dq_1^2dq_2^2d\phi.
\end{equation}

Thus, the function $F$ is determined by 15 parameters listed in 
Table~\ref{tab:fit}. Parameters $\alpha$ and $\Im_{\eta S}$ are fixed in the
$\eta_c$ case, so they do not have a statistical error. In case the model
error is not quoted, it is negligibly small compared to the statistical one.
\begin{table}[!h]
\begin{center}
\caption{Fit results.}
\begin{tabular}{|c||c|c|} 
\hline
Parameter & $B^{\pm}\to K^{\pm}\eta_c$,
$\eta_c\to (K_S K\pi)^0$ &$ B^{\pm}\to K^{\pm}\eta_c(2S)$,
$\eta_c(2S)\to (K_S K\pi)^0$\\
\hline\hline
$N$ & $920\pm 50(stat)\pm 170(model)$ & $128^{+83}_{-58}(stat+model)$\\
$\alpha$ & $4.8\pm 2.3(model)$ & $8.4^{+1.1}_{-3.3}(stat+model)$\\
$\beta$ & $5.3\pm 1.6(stat)\pm 2.1(model)$ & $3.7^{+3.8}_{-2.1}(stat+model)$\\
$\gamma$ & $7.9\pm 1.4(stat)$ & $6.2^{+1.5}_{-1.8}(stat+model)$\\
$\xi_{\eta S}$ & $0.37\pm 0.24(stat)\pm 0.18(model)$ & $0.84^{+0.16}_{-0.45}(stat+model)$\\
$\Im_{\eta S}$ & $-0.06\pm 0.80(model)$ & $-0.39^{+1.12}_{-0.53}(stat+model)$\\
$\xi_{\eta P}$ & $0.48\pm 0.22(stat)\pm 0.19(model)$ & $0.51^{+0.49}_{-0.51}(stat+model)$\\
$\theta_{\eta P}$, rad & $0.08\pm 0.65(stat)$ & $-0.24^{+3.14}_{-3.14}(stat+model)$\\
$\xi_{\eta D}$ & $0.22\pm 0.16(stat)\pm 0.02(model)$ & $0.81^{+0.19}_{-0.62}(stat+model)$\\
$\theta_{\eta D}$, rad & $0.64\pm 0.80(stat)$ & $0.58^{+0.85}_{-1.10}(stat+model)$\\
$\Pi_{SP}$ & $-0.74\pm 0.46(stat)\pm 0.23(model)$ & $-1^{+2}_{-0}(stat+model)$\\
$\Pi_{SD}$ & $-0.11\pm 0.34(stat)\pm 0.42(model)$ & $-1^{+2}_{-0}(stat+model)$\\
$\Pi_{PD}$ & $0.24\pm 0.77(stat)\pm 0.50(model)$ & $-0.95^{+1.95}_{-0.05}(stat+model)$\\
$M$, MeV/$c^2$ & $2985.4\pm 1.5(stat)$ & $3636.1^{+3.9}_{-4.2}(stat+model)$\\
$\Gamma$, MeV/$c^2$ & $35.1\pm 3.1(stat)$ & $6.6^{+8.4}_{-5.1}(stat+model)$\\ 
\hline
\end{tabular}
\label{tab:fit}
\end{center}
\end{table}

Function $F(s,x)$ can be represented as a rational 
function of $s$ and $x$:
\begin{equation}
F(s,x)=\frac{1+\varepsilon_1 x+\varepsilon_2 x^2}{(s-M^2)^2+M^2\Gamma^2}\sum_{i=0}^{2}\sum_{j=0}^{4}C_{ij}s^ix^j,
\end{equation}
where
\begin{enumerate}
\item $C_{24}=\frac{45}{8}\gamma^2$,
\item $C_{23}=\frac{3}{2}\sqrt{15}\beta\gamma\Pi_{PD}$,
\item $C_{22}=\frac{3}{2}\beta^2-\frac{15}{4}\gamma^2+\frac{3}{2}\sqrt{5}\alpha\gamma\Pi_{SD}$,
\item $C_{21}=\sqrt{3}\alpha\beta\Pi_{SP}-\frac{\sqrt{15}}{2}\beta\gamma\Pi_{PD}$,
\item $C_{20}=\frac{1}{2}\alpha^2+\frac{5}{8}\gamma^2-\frac{\sqrt{5}}{2}\alpha\gamma\Pi_{SD}$,
\item $C_{14}=-2M^2C_{24}$,
\item $C_{13}=-2M^2C_{23}$,
\item $C_{12}=\frac{3}{2}\sqrt{5}\sqrt{N}\gamma\Re_{\eta D}-2M^2C_{22}$,
\item $C_{11}=\sqrt{3}\sqrt{N}\beta\Re_{\eta P}-2M^2C_{21}$,
\item $C_{10}=\sqrt{N}\alpha\Re_{\eta S}-\frac{\sqrt{5}}{2}\sqrt{N}\gamma\Re_{\eta D}-2M^2C_{20}$,
\item $C_{04}=M^2(M^2+\Gamma^2)C_{24}$,
\item $C_{03}=M^2(M^2+\Gamma^2)C_{23}$,
\item $C_{02}=\frac{3}{2}\sqrt{5}\sqrt{N}\gamma(M\Gamma\Im_{\eta D}-M^2\Re_{\eta D})+M^2(M^2+\Gamma^2)C_{22}$,
\item $C_{01}=\sqrt{3}\sqrt{N}\beta(M\Gamma\Im_{\eta P}-M^2\Re_{\eta P})+M^2(M^2+\Gamma^2)C_{21}$,
\item $C_{00}=\frac{1}{2}N+\sqrt{N}\alpha(M\Gamma\Im_{\eta S}-M^2\Re_{\eta S})-\frac{\sqrt{5}}{2}\sqrt{N}\gamma(M\Gamma\Im_{\eta D}-M^2\Re_{\eta D})+M^2(M^2+\Gamma^2)C_{20}$.
\end{enumerate}

\end{document}